\documentclass[superscriptaddress,nobibnotes,amsmath,amssymb,notitlepage,twocolumn,aps,pra,longbibliography,10pt,floatfix]{revtex4-2}

\usepackage{bm,braket}
\usepackage[toc,page]{appendix}
\usepackage{comment}
\usepackage{amsfonts}

\usepackage{xcolor}
\usepackage{bbm}
\usepackage{dcolumn} % Align table columns on decimal point
\usepackage{siunitx}
\usepackage{enumitem}  % to use (i) etc for enumerate lists
\usepackage{graphicx,color,hyperref}
\usepackage[caption=false]{subfig}
\hypersetup{colorlinks=true, linkcolor=blue, citecolor=blue, urlcolor=blue} 

\graphicspath{{./figures/}}
\newcommand{\beq}[1]{\begin{equation}\label{#1}}
\newcommand{\eep}{\;.\end{equation}}
\newcommand{\eec}{\;,\end{equation}}
\newcommand{\eeq}{\end{equation}}

\newcommand*\dd{\mathop{}\!\mathrm{d}} %differential d

\DeclareMathOperator{\tr}{tr}

\newcommand{\lb}{\left(}
\newcommand{\rb}{\right)}

\renewcommand{\a}{\alpha}
\renewcommand{\b}{\beta}
\newcommand{\g}{\gamma}
\renewcommand{\d}{\delta}

\newcommand{\ps}{\psi}
\newcommand{\om}{\omega}

\newcommand{\D}{\Delta}
\newcommand{\G}{\Gamma}

\newcommand{\Om}{\Omega}

\DeclareMathAlphabet{\mathcal}{OMS}{cmsy}{m}{n} % Changes font for mathcal but leaves the rest of the math fonts in Times.

\renewcommand{\vec}[1]{{\bf #1}}

\newcommand{\dv}{\vec{d}}
\newcommand{\kv}{\vec{k}} % moved to main
\newcommand{\R}{\vec{R}} 
\newcommand{\rv}{\vec{r}}

\newcommand{\q}{\vec{q}}

\newcommand{\uv}{\vec{u}} % moved to main

\makeatletter
\renewcommand*{\fnum@figure}{{\normalfont\bfseries \figurename~\thefigure}}
\makeatother

\allowdisplaybreaks

\definecolor{orange}{rgb}{1,0.5,0}

\DeclareMathAlphabet{\mathcal}{OMS}{cmsy}{m}{n} % Changes font for mathcal but leaves the rest of the math fonts in Times.

\renewcommand{\Re}{\mathfrak{Re}~}
\renewcommand{\Im}{\mathfrak{Im}~}

\makeatletter
\newcommand{\specificthanks}[1]{\@fnsymbol{#1}}% Inserts a specific \thanks symbol
\makeatother

\newcommand{\Tr}{\operatorname{Tr}}

\newcommand{\Ov}{\vec{0}}
\newcommand{\Gv}{\vec{G}}
\newcommand{\vv}{\vec{v}}
\newcommand{\ptl}{\partial}
\newcommand{\qgt}{\mathcal{Q}}
\newcommand{\Hm}{\mathrm{H}}
\newcommand{\Am}{\mathrm{A}}
\newcommand{\Um}{\mathrm{U}}
\newcommand{\Vm}{\mathrm{V}}
\newcommand{\Pm}{\mathrm{P}}
\newcommand{\Qm}{\mathrm{Q}}
\newcommand{\Dm}{\mathrm{D}}

\newcommand{\sumNM}{\sum_{\substack{n\, \mathrm{occ} \\ 
                    m\,\mathrm{unocc}}}}

\usepackage[section, above, below]{placeins}

\graphicspath{{./figures/Conceptual}}

\begin{document}

\preprint{APS/123-QED}

\title{Optimally embedded tight binding for reproducing geometry dependent observables}% Force line breaks with \\

\author{Jonas J. Telle}
 \email{jonas.telle@gmail.com}
 \affiliation{Department of Physics, University of Oslo, N-0316 Oslo, Norway}
\author{Gunnar F. Lange}
 \email{gunnalan@uio.no}
\affiliation{Department of Physics, University of Oslo, N-0316 Oslo, Norway}
\affiliation{Centre for Materials Science and Nanotechnology, University of Oslo, N-0316 Oslo, Norway}

\date{\today}

\begin{abstract}
 In tight-binding models, the position operator is reduced to intra-cell orbital positions (embeddings). While accurately reproducing band structures, such models often fail for geometry dependent responses depending on the position operator. To address this, we investigate the role of these embeddings and introduce the general framework of optimally embedded tight binding. Treating the embeddings as geometric tuning parameters to be fixed against a reference response (obtained from ab-initio computation or experiment), we obtain tight-binding models of GaAs and CdS which quantitatively reproduce non-linear optical responses at no cost to band structure accuracy. The optimal embeddings are determined efficiently using position derivatives obtained from decomposing tight-binding observables into a geometry independent and dependent part, and explicit derivatives are provided for key quantities such as the quantum geometric tensor. The decomposition reveals where geometric effects dominate, and we show in both toy models and in the Chern insulator V$_2$O$_3$ how geometry can dramatically alter the local metric trace. Our results highlight that orbital embeddings should be treated as a genuine model parameter which should be explicitly fixed against physical data to get accurate minimal models.
\end{abstract}

\maketitle

\section{\label{sec:introduction}Introduction}

\begin{figure}[h!]
    \centering
    \def\svgwidth{\linewidth}
    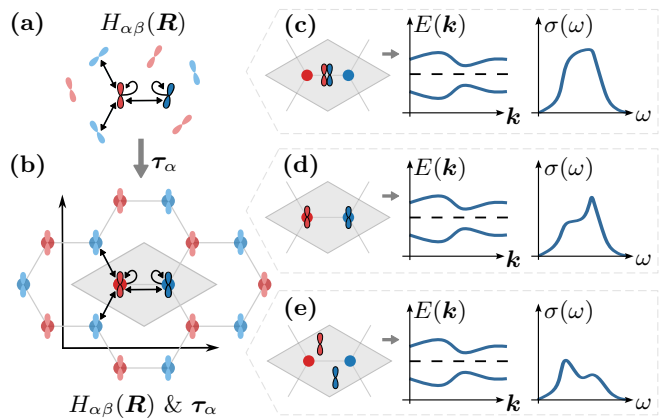
    \caption{Schematic of main results. \textbf{(a)} A tight-binding model can be viewed as a weighted graph, where hoppings and on-site terms are represented by edges and loops. \textbf{(b)} Specifying the orbital positions $\boldsymbol{\tau}_\a$ amounts to embedding this graph in real space. \textbf{(c)--(e)} Changing the orbital positions leaves geometry independent quantities such as band structures $E(\kv)$ unchanged, but can dramatically alter geometry dependent quantities such as optical responses $\sigma(\omega)$. Thus, the orbital positions can be seen as a tuning parameter which can be utilised to fit geometric quantities.}
    \label{fig:conceptual_fig}
\end{figure}

The tight-binding method is perhaps the most widely applied technique for computing electronic properties of solid-state systems. By expanding energy eigenstates as a linear combination of basis orbitals, one can reduce wave functions in (and operators on) Hilbert space to coefficient vectors in (and matrices on) a smaller projected space.
In its simplest form, a tight-binding model can be specified by providing an effective real-space Hamiltonian matrix $H_{\alpha\beta}(\boldsymbol{R})$ in terms of on-site energies $\epsilon_{\alpha}$
and hopping terms $t_{\alpha\beta}(\boldsymbol{R})$ for each unit cell $\R$, as visualised as an abstract graph in Fig.~\ref{fig:conceptual_fig}(a). The model parameters can be fitted directly to reproduce a band structure known from experiment or ab-initio computation, circumventing the need for constructing a real-space representation for the basis orbitals at all.

Several quantities, such as band structures and topological invariants, are completely determined by this simple model. However, many fundamental observables such as optical responses \cite{ahn_riemannian_2022}, polarization \cite{vanderbilt_berry_2018}, magnetization \cite{Magnetization_quantum_geometry} as well as superconductivity \cite{peotta_superfluidity_2015,huhtinen_revisiting_2022}, depend explicitly on the position operator $\boldsymbol{\hat{r}}$. We refer to such properties as geometry \textit{dependent}, in contrast to the geometry \textit{independent} properties which do not depend on the position operator \cite{simon_contrasting_2020}. 

A first step towards incorporating geometric information in tight binding is to specify only the \textit{diagonal} elements  $\boldsymbol{\tau}_{\alpha}$ of the position operator matrix. These positions give a spatial embedding \cite{1D_embedding,lim_geometry_2015,simon_contrasting_2020} of the graph representing the model, as shown in Fig.~\ref{fig:conceptual_fig}(b). Physically, this approximation takes the orbitals to be nearly point-like, and is theoretically very attractive as it ensures clean electromagnetic coupling and simplifies symmetry operators. 

In analytical models, the orbital embeddings are typically chosen to sit on atomic sites or fixed by symmetries. For larger models, the Wyckoff positions are not unique, and a more principled approach used in computational work proceeds through Wannierisation (by e.g. \texttt{Wannier90} \cite{mostofi_updated_2014}) after a first-principles computation, creating a real-space representation of the orbitals in terms of maximally localised Wannier functions (MLWFs). The Wannier charge centers (WCCs) are then typically used for the tight-binding positions.

In the computational approach, the predicted band structure matches the first-principles computation and can be easily verified against experimental data. However, as we discuss, when the orbital centres are chosen to satisfy maximal localization of Wannier functions, they are not inherently physical. As such, one might be motivated to consider models with the same hoppings and on-site energies, but different orbital positions. Such models will have the same geometry independent quantities, and in particular the same band structure, but differ in their geometry dependent quantities, as illustrated in \mbox{Fig.~\ref{fig:conceptual_fig}(c)--(e)}.

We will refer to any model fully specified by $\{t_{\alpha\beta}, \epsilon_{\alpha}, \boldsymbol{\tau}_{\alpha}\}$ (and its real-space lattice vectors) as a tight-binding model \footnote{This is sometimes called \textit{diagonal} tight binding.}. Notice that there are many different tight-binding models for a given band structure, differing in the choice of $\boldsymbol{\tau}_{\alpha}$. Interpreting $\boldsymbol{\tau}_{\alpha}$ not as Wannier centres, but as a tuning parameter that changes the geometry dependent properties of the tight-binding models without affecting the geometry independent properties, allows us to optimize the orbital positions to fit geometric quantities.

This work demonstrates that this tuning allows us to reproduce geometric observables to significantly better precision than has previously been achieved with tight-binding models \cite{wang_first-principles_2017, ibanez-azpiroz_ab_2018, ghosh_choosing_2025}. Specifically, we show that fitting general geometric objects, such as the Abelian quantum geometric tensor (QGT), can give significant improvements on optical responses, and that if a relevant response is available in a more accurate method such as density-functional theory (DFT), an optimal fit can be obtained with very small additional cost. 
As every tight-binding model has comparable complexity, we can therefore generate models that preserve the simplicity of tight binding while simultaneously improving the accuracy of the model's predictions on geometry dependent observables.

Concretely, our method proceeds by running an ab-initio electronic calculation, which we Wannierise to fix the on-site and hopping elements of our model. Subsequently, we use the method of Wannier interpolation to obtain precise estimates of a geometric quantity of interest, to which we then fit the orbital positions. This gives tight-binding models which by design reproduce specific geometric quantities well, without sacrificing any accuracy in non-geometric quantities such as band structures.

It is worth noting that optimally embedded tight binding in no way relies on computing real-space representations of Wannier functions. For our purposes, they serve as a particularly convenient way of obtaining an effective Hamiltonian and interpolated reference responses, but given a real-space tight-binding Hamiltonian and a reference geometric response one can always find optimal embeddings to best capture that response.

In developing the theory facilitating optimal embeddings, we obtain useful tools for studying the isolated effect of geometry in systems adequately modelled by tight binding. Understanding where geometric effects are prominent is of practical importance for assessing when common approximations are appropriate and for controlling errors. Furthermore, we want to know how geometry might qualitatively alter the shape of central objects such as the quantum metric, as this might suggest interesting physics in otherwise unremarkable physical platforms. We therefore employ our theory to show that displacing the orbital embeddings can dramatically alter the qualitative shape of the metric trace in the Chern insulator V$_2$O$_3$ and in toy models.

The paper is structured as follows. In Sec.~\ref{sec:background}, we give an overview of tight binding and Wannier functions, introducing useful notation and terminology to clear up common confusions. We particularly discuss the diagonal tight-binding assumption and the two common conventions for how the embeddings enter tight-binding models. Wannier interpolation is introduced as a baseline against which we will evaluate tight binding. Readers familiar with these concepts may skip this section and refer back to equations only as referenced. 
In Sec.~\ref{sec:theory}, we introduce geometry decomposition and position derivatives, useful in superconductivity \cite{huhtinen_revisiting_2022} for finding the minimal metric trace embedding quickly for large models and for studying geometry dependence, but here most importantly for obtaining optimal embeddings. We apply this in Sec.~\ref{sec:optimal}, showing and discussing results of choosing optimal embeddings for GaAs and CdS on first and second-order optical responses. As a second application, we study qualitative effects of geometry in Sec.~\ref{sec:geometry}, suggesting where we expect geometric effects to be strong before presenting and discussing results on the geometry dependence of the metric trace for the Chern insulator V$_2$O$_3$ as well as standard toy models. Finally, we conclude in Sec.~\ref{sec:conclusion}, providing some suggestions for future work.

\section{Background\label{sec:background}}

\subsection{Tight binding}

A tight-binding model can be viewed as a weighted graph, as illustrated in Fig.~\ref{fig:conceptual_fig}(a), where the nodes are electron sites $\ket{\a \R}$ and the edges encode the Hamiltonian matrix elements $H_{\a\b}(\R) = \braket{\a\Ov | \hat H | \b\R}$. By Fourier transformation, we reduce the infinite set of periodic images of the site $\ket{\a\Ov}$ to a single $\kv$-dependent Bloch-like basis state
\begin{equation} \label{eq:bloch_basis}
    \ket{\tilde \psi^\a_\kv} = \frac{1}{\sqrt{N}}\sum_\R e^{i\kv\cdot \R} \ket{\a\R}.
\end{equation}
Here, $N$ is the number of real-space cells, which is taken to infinity in the thermodynamic limit. 

\subsubsection{The Hamiltonian}
Expanding the Hamiltonian in this Bloch-like basis, we get the $\kv$-dependent finite-dimensional Hamiltonian matrix
\begin{align} \label{eq:H_k-periodic}
    H^{\a\b}_\kv = \braket{\tilde \psi^\a_\kv | \hat H |\tilde \psi^\b_\kv}
    % \\ = \frac{1}{N}\sum_{\R',\R''} e^{i\kv\cdot(\R'' - \R')}\braket{n\R' | \hat H |\ m\R''} \\
    = \sum_\R e^{i\kv \cdot \R} H_{\a\b}(\R)
\end{align}
which we diagonalise to obtain the band structure $E_{n\kv}$ and a unitary matrix $\Um(\kv)$ whose columns contain the coefficients of the Bloch eigenstates $\ket{\psi_{n\kv}}$ in the Bloch-like basis:
\begin{equation} \label{eq:ham_basis}
    \ket{\psi_{n\kv}} = \sum_a\ket{\tilde \psi^a_\kv}U_{an}(\kv).
\end{equation}
This unitary matrix is not uniquely defined, giving rise to a gauge freedom: We may exchange columns, multiply single columns by a phase %$e^{-i\varphi_n(\kv)}$
or, more generally, mix states that are degenerate in energy with unitary rotations. While it is common to call \textit{any} unitary rotation of the Bloch states a multi-band gauge choice, we will resist this temptation and reserve the word for the ambiguity within matrices diagonalising the Hamiltonian, corresponding to the arbitrary choice of representative for each eigenspace. 

Finally, note that the band structure $E_{n\kv}$ was obtained with no reference to the position operator and is therefore geometry independent.

\subsubsection{The position operator}

Since the position operator does not affect band structures, it is sometimes overlooked. However, interest in topology has highlighted its importance, since it enters the Berry connection through the cell-periodic Bloch states 
\begin{equation} \label{eq:cell_periodic_Bloch}
    \ket{u_{n\kv}} = e^{-i\kv\cdot\hat\rv}\ket{\psi_{n\kv}}.  
\end{equation}

Note that we need cell-periodicity to make sense of derivatives in $\kv$-space: From Eqs. \eqref{eq:bloch_basis} and \eqref{eq:ham_basis}, it is clear that the states $\ket{\psi_{n\kv}}$ are eigenstates of unitary lattice translations $\mathrm T_\R$ with eigenvalues $e^{-i\kv\cdot\R}$. They are therefore orthogonal for all small separations $\d\kv$, so that $k_j$-derivatives $\ptl_j\ket{\psi_{n\kv}}$ are ill-defined. The exponentiated position operator in \eqref{eq:cell_periodic_Bloch} exchanges $\kv$-periodicity for cell-periodicity, allowing us to define a normalised inner product
\begin{align*}
    \braket{u_{n\kv}|u_{n(\kv+\d\kv)}} &\equiv \frac{1}{N} \int \dd{\rv} \, u^*_{n\kv}(\rv) u_{n(\kv+\d\kv)}(\rv)\\
    &= \int_{\text{cell}} \dd{\rv} \, u^*_{n\kv}u_{n(\kv+\d\kv)},
\end{align*}
where the first integral is over all of space. This is now well-behaved and defines a derivative as usual \cite{vanderbilt_berry_2018}.

It is convenient to define corresponding cell-periodic basis states
$\ket{\tilde u^\a_\kv} = e^{-i\kv\cdot \hat \rv}\ket{\tilde \psi^\a_\kv}$
and compute the orbital representation of the Berry connection, defined as
\begin{align}
    \tilde A^{\a\b}_j \equiv \braket{\tilde u^\a_\kv| i\ptl_j | \tilde u^\b_\kv}
    =  \sum_\R e^{i\kv\cdot\R} \braket{\a\Ov|\hat r_j|\b\R}, \label{eq:WanA} 
\end{align}
where the final equality was obtained by expanding the basis states with \eqref{eq:bloch_basis} and noting by periodicity that the position matrix elements have the property $\braket{\a\R' | \lb \hat \rv - \R' \rb|\b(\R+\R')} = \braket{\a\Ov | \hat \rv |\b\R}$. 
The 
%standard Hamiltonian representation of the 
Berry connection can then be expressed with the chain rule as
\begin{align}
    A^{nm}_j &\equiv \braket{u_{n\kv}|i\ptl_j|u_{m\kv}} = \sum_{a,\,b} U^\dagger_{na}\bra{\tilde u^a_\kv} i\ptl_j\lb \ket{\tilde u^b_\kv}U_{b m} \rb \notag \\
    &= [\Um^\dagger\tilde \Am_j\Um + i\Um^\dagger\ptl_j\Um]^{nm},\label{eq:HamA}
\end{align}
where the final term in the last line is a geometry independent contribution, since it depends only on the derivatives of the column vectors $\vec{u}_n$ of $\Um$ obtained by diagonalising the Hamiltonian.

While the degenerate terms ($E_n=E_m$) are not gauge covariant \cite{wang_ab_2006}, the non-degenerate terms of $\Um^\dagger \ptl_j\Um$, with $n,m$ such that $E_n \neq E_m$, are fixed by considering
\begin{align*}
    \ptl_j \Hm_\kv\vec{u}_n &= \ptl_j(\Hm_\kv\vec{u}_n) - \Hm_\kv \ptl_j \vec{u}_n \\
    &= \ptl_j E_n\vec{u}_n + (E_n-\Hm_\kv)\ptl_j\vec{u}_n
\end{align*}
and multiplying from the left by $\vec{u}_m^\dagger$ to get
\begin{equation} \label{eq:A_Kubo}
    \frac{\uv_m^\dagger \ptl_j \Hm_\kv \uv_n}{E_n-E_m} = \vec{u}_m^\dagger \ptl_j \vec{u}_n.
\end{equation}

\subsubsection{The tight-binding assumption\label{sssec:TBA}}

In practice, geometry is typically included in a tight-binding model by assigning a real space position $\tau^\a_j$ to each home-cell orbital. These positions are sometimes called orbital embeddings, since they determine the placement of the nodes of the hopping graph in a coordinate plane, as illustrated in Fig.~ \ref{fig:conceptual_fig}(b).

Within this approach, one only accounts for diagonal position matrix elements
\[
    \braket{\a\Ov | \hat \rv_j | \b\R} \simeq \tau_j^\a \delta_{\a\b}\,\delta_{\Ov\R}. \tag{TBA}
\]
We refer to this as the (diagonal) tight-binding assumption (TBA) \cite{ibanez-azpiroz_ab_2018}. Whenever we refer to tight binding, it will always be in the sense of diagonal tight binding.

With the TBA, the exponentiated position operator takes a particularly simple form, and it is tempting to define a new Bloch-like basis
\begin{align}
    \ket{\check\psi_\kv^\a} &= \frac{1}{\sqrt{N}}\sum_\R e^{i\kv\cdot\hat \rv}\ket{\a\R} \notag\\
    &= \frac{1}{\sqrt{N}}\sum_\R e^{i\kv\cdot(\R+\bm \tau_\a)}\ket{\a\R} =e^{i\kv\cdot\bm\tau_\a}\ket{\tilde \psi_\kv^\a}. \label{eq:CI}
\end{align}
This choice of Bloch-like basis, where one includes the orbital embeddings in the Fourier kernel,  is often denoted Convention I \cite{vanderbilt_berry_2018}, but we will simply call it the tight-binding convention (TBC). The last equality above highlights that the TBC corresponds to a diagonal unitary transformation of the standard Bloch-like basis. As such, the TBC Hamiltonian is
\begin{equation}
    \check H^{\a\b}_\kv = e^{-i\kv\cdot\bm\tau^\a} H^{\a\b}_\kv e^{i\kv\cdot\bm\tau^\b}, \label{eq:HamCI}
\end{equation}
and the diagonalising matrix therefore becomes
\begin{equation} \label{eq:V}
    V_{\a n} = e^{-i\kv\cdot\bm\tau^\a}U_{\a n},
\end{equation}
where $\Um$ is the unitary rotation from before.

The advantage of the TBC is that the Berry connection takes a particularly simple form: In the TBA, the orbital representation of the Berry connection is diagonal $\tilde A^{\a\b}_j = \tau^\a_j\delta_{\a\b}$, as is immediate from \eqref{eq:WanA}. We can then write the Berry connection as
\[
    \Am_j = i\Vm^\dagger\ptl_j \Vm,
\]
where $\Vm$ is the diagonalising matrix for the TBC Hamiltonian in \eqref{eq:V},
as is seen by carrying out the derivative with the product rule and comparing with \eqref{eq:HamA}. Denoting the columns of $\Vm$ by $\vv_n$, we see that
\[
    \Am_j^{nm} = \braket{u_{n\kv} |i\ptl_j|u_{m\kv}} = \vv_n^\dagger i\ptl_j \vv_m.
\]
Hence, the coefficients $\vv_n$ obtained by diagonalising the TBC Hamiltonian contain all the geometric information of the state  $\ket{u_{n\kv}}$, assuming the TBA: Expressions in the TBC are simply obtained from general expressions, with derivatives written in terms of the Berry connection, by replacing operators with the corresponding matrices and the states $\ket{u_{n\kv}}$ by the corresponding vectors $\vv_n$. In this sense, the geometry is already fully incorporated in the TBC Hamiltonian, which proves convenient for tight-binding computation since one does not have to keep track of the geometric correction terms. Furthermore, the TBC Hamiltonian explicitly incorporates the real-space symmetries of the system, as discussed in \cite{dobardzic_generalized_2015}.
It is hence an obvious choice when studying for example symmetry-protected topology within the TBA.

In this work, however, we are interested in making all geometry dependence explicit. Hence, we will stick to the basis in \eqref{eq:bloch_basis}, often denoted Convention II, but which we shall label more descriptively as the $\kv$-periodic convention (KPC), where the Hamiltonian knows nothing about the orbital embeddings. As the name suggests, this convention makes the Hamiltonian periodic in $\kv$-space. 

We emphasize that there is no physical difference between the conventions (they correspond only to two different choices of Bloch-like basis). When using the KPC, one must remember to add the geometric corrections as additional terms. Without these, the KPC has no geometry, in the sense that the projected position operator is zero, and the computed observables would correspond to a tight-binding model in the TBC with every home-cell orbital placed on the origin, as shown in Fig.~\ref{fig:conceptual_fig}(c). This observation allows us to decompose any tight-binding observable into geometry dependent and independent terms, as will be explored in section \ref{ssec:decomposition}.

\subsection{\label{ssec:wannier}Wannier functions}

Wannier functions provide both a solid theoretical justification for the tight-binding framework and a practical way to construct tight-binding models for real materials \cite{MLWF_review}. In particular, they allow us to compute the Hamiltonian and position matrix used in the previous section, and more generally, to densely interpolate quantities in $\kv$-space computed on a coarse ab initio grid. The discussion here mirrors the previous section, but in reverse, starting from single particle wave functions in $\kv$-space to arrive at a tight-binding model \cite{wang_ab_2006}.

\subsubsection{Wannier interpolation}

For a system governed by an effective single-particle periodic Hamiltonian, we have by Bloch's theorem cell-periodic states $\ket{u_{n\kv}} = e^{-i\kv\cdot\hat\rv}\ket{\ps_{n\kv}}$ labelled by translation phases $\kv$ and a band index $n$. For real materials, one usually obtains these from a first-principles calculation (e.g. plane-wave DFT) on a coarse grid, whose discrete points will be labelled by $\q$, in the dense $\kv$-space.

We want to capture the physics within some energy cut-off around the Fermi level. Suppose there are $N$ bands which are at least partially within this cut-off. For simplicity, we assume that this group of bands never intersects bands outside the group (we discuss the general case in Sec.~\ref{sssec:approximations}) so that the projector
\begin{equation} \label{eq:projector}
    \hat P_{\q} = \sum_{n=1}^N \ket{u_{n\q}}\bra{u_{n\q}}
\end{equation}
is independent of the choice of representative for the bands (this would otherwise not be true at degeneracies with bands outside the group).

We can choose any orthonormal basis for this subspace with a (suggestively named) unitary rotation $\Um^\dagger (\q)$,
\begin{equation} \label{eq:wannier_basis}
        \ket{\tilde u^\a_\q} = \sum_n\ket{u_{n\q}}U_{n\a}^\dagger \quad \text{with} \quad \ket{\tilde \psi^\a_\kv} = e^{i\q\cdot\hat\rv}\ket{\tilde u^\a_\q},
\end{equation} 
and compute matrix representations of projected operators, such as
\[
    H^{\a\b}_\q 
    %= \braket{\tilde u^\a_\q|e^{-i\q\cdot\hat\rv}\hat H e^{i\q\cdot\hat\rv}|\tilde u^\b_\q} 
    = \braket{\tilde \psi^\a_\q|\hat H|\tilde \psi^\b_\q} \quad \text{and} \quad  \tilde A^{\a\b}_j = \braket{\tilde u^\a_\q | i\ptl_j |\tilde u^\b_\q}.
\] 

The latter corresponds to the position operator per \eqref{eq:WanA}, and the derivative is typically computed by finite differences on the coarse $\q$-grid, though an exact $\q$-local expression for a particular basis choice was recently presented in \cite{cole_exact_2026}.

We Fourier transform to get orthonormal Wannier states
\[
    \ket{\a\R} = \int_\q e^{-i\q\cdot\R} \ket{\tilde \psi_{\a\q}},
\]
where we introduced short-hand notation for the Brillouin zone (BZ) average (which is discretized in practice, and here given in reduced coordinates):
\[
\int_\q \equiv \int_{[0,2\pi)^d} \frac{\dd^d q}{(2\pi)^d}.
\]
The Wannier states are periodic images of home-cell representatives and can thus be represented by a finite set of Wannier functions $w_\a(\rv-\R) = \braket{\rv|\a\R}$. 

Furthermore, since the Fourier transformation is invertible, the band structure within the projected window of interest is reproduced exactly by the hoppings
\begin{align} \label{eq:hoppings}
    H_{\a\b}(\R) = \braket{\a\Ov | \hat H | \b\R } = \int_\q e^{-i\q\cdot\R}H^{\a\b}_\q. 
\end{align}
Similarly, we obtain the real-space position matrix by inverting \eqref{eq:WanA} \cite{wang_ab_2006}. 

Both of these matrices are available directly from \texttt{Wannier90} and make up all the ingredients used in the previous subsection. Hence, we can hereafter throw away all information about the complicated Bloch states and Wannier functions to get a cheap model of the physics near the Fermi level. This is enormously useful, as we can now interpolate the coarse $\q$-grid from our first principles calculation by choosing a much denser grid for the Fourier transformation back to reciprocal space. This allows us to accurately perform integrals of quantities with very sharp features in $\kv$-space, which is in particular important for accurately computing optical responses.

\subsubsection{Localisation}

The localisation of Wannier functions is of great theoretical importance and underlies standard approximations like the Peierls substitution and the TBA. Formally, it was shown in \cite{brouder_exponential_2007} that exponentially localised Wannier functions spanning a projected space exist (in dimensions $\leq 3$) if (and only if) the Chern number(s) associated with that projected space all vanish, 
% or more precisely, of the U(N) bundle over the brillouin zone corresponding to P_k
and that this choice corresponds to making the Bloch like basis analytic in $\kv$. This is a rather general result, and even if a certain occupied band space has non-trivial topology, we can still hope to choose a larger projected space that is trivial by including more bands, which would then still give a good tight-binding model for the system.

In practice, then, we can utilize the free choice of basis in \eqref{eq:wannier_basis} to tune the Bloch basis to give \textit{maximally} localized Wannier functions (MLWFs) \cite{marzari_maximally_1997}. This is typically done with \texttt{Wannier90} \cite{mostofi_updated_2014}. For a topologically trivial projected space, this basis choice ensures that matrix elements between Wannier functions fall off quickly with $|\R|$, so that we can neglect long-range hoppings. Furthermore, localisation ensures that the derivative
\[
    \ptl_j H_\kv^{\a\b} = \sum_\R iR_j e^{i\kv\cdot \R} H_{\a\b}(\R)
\] 
is convergent, and exponential localisation ensures that all derivatives exist.

\subsubsection{\label{sssec:approximations}Examining assumptions}

It is worth being explicit about the assumptions made in the construction of a tight-binding model making it less accurate than a full DFT description. 

First, the previous section shows that long range real space matrix elements will be negligible. Striking a balance between minimality and accuracy is an interesting practical problem, but we could in principle include enough terms that this will never be the limiting assumption.

Second, the projection operator for the projected space clearly takes a trivial form in matrix representation, $\braket{\tilde u^\a_\q |\hat P_\q |\tilde u^\b_\q} = \delta^{\a\b}$. This corresponds to an energy cut-off for sums over intermediate states, in the sense that it defines the unoccupied projection $\Qm = 1-\Pm$, where $\Pm$ denotes the occupied space projection matrix and we write $1$ for the identity matrix $\d^{\a\b}$. 

A trivial consequence of throwing away intermediate states in this way is that our tight-binding model knows nothing of neither the topology nor the geometry of the full projected space. In particular, if one chooses a projected subspace with a non-zero Chern number, one will not be able to compute this from within the resulting model (though one might notice poor localisation of the Wannier functions).

Ideally, then, we would want the projector $\hat P_\q$ to be constant in $\q$ inside the large Hilbert space. In this case, a derivative of a state inside the projected space does not leak into the neglected intermediate states at all, so that there is no geometric information loss out of the projection. While this will not generally be possible, we can quantify the geometric information loss as the square norm of the leakage out of the projected space from infinitesimal displacements in $\kv$ by the Fubini-Study metric trace for the projected space, as
\begin{equation} \label{eq:tr_g_leakage}
    \sum_{n,\,j} \| (1-\hat P_\q) \ptl_j\ket{u_{n\q}}\|^2 = \sum_{n, j} \qgt^{nn}_{jj}(\q) = \tr g(\q).
\end{equation}
Here, we used the definition of the multi-band QGT for a subspace with projector $\Pm$,
\[
    \qgt_{ij}^{nm} = \braket{\ptl_i u_{n\kv} | 1 - \Pm | \ptl_j u_{m\kv}}.
\]
Taking the band trace gives the Abelian QGT, which has the quantum metric $g_{ij}$ as its symmetric real part and the Berry curvature $\Om_{ij}$ is proportional to its anti-symmetric imaginary part.

We hence want to chose the projected space to minimize the BZ average of this metric trace, known as the quantum weight \cite{onishi_quantum_2025, marzari_maximally_1997}. Beyond choosing which bands to project onto, corresponding to the choice of energy cut-off, this is done by disentanglement: In Sec.~\ref{ssec:wannier}, we assumed that the energy cut-off determined an isolated group of bands. This assumption is not needed in practice, as the situation with band crossings is reconciled by standard disentanglement procedures \cite{WF_disentaglement} implemented in e.g. \texttt{Wannier90}, mixing bands above the strict energy cut-off to minimise the projected space quantum weight. 

The band structure for the projected model remains exact below the cut-off, as exemplified in Fig.~\ref{fig:CdS_response}(g). 

We should note that frozen window disentanglement is slightly symmetry breaking. When symmetry is vital, such as for crystalline topological insulators, spin-topology or analytical models, symmetry adapted Wannier functions may be used \cite{Symmetry_adapted_WF}, though this might come at the cost of localisation or band structure accuracy. Then, the WCCs are forced to occupy symmetry positions, and the TBA with WCCs as orbital embeddings will not break symmetry further. For most practical applications, however, the symmetry breaking is sufficiently weak as to be of little importance, and we verify in our applications that symmetry-disallowed observables are suppressed. 

Barring these details on symmetry breaking from disentanglement, the finite space approximation is again mostly a practical limitation: One could always include more intermediate states to improve the approximation.

Assuming the approximations above have been made, taking point-like orbitals is the most natural next reduction in light of localisation. It also proves exceptionally useful, reducing algebra, clarifying symmetries and dramatically simplifying electromagnetic coupling.

However, while the total space projector looks trivial from inside the tight-binding model, it determines the structure of the projected operators in interesting ways. In particular, one might think that the TBA simply amounts to choosing a basis diagonalising the position operator. However, there is no such choice, since the projected position components generally do not commute \footnote{The condition that they do commute is exceptionally strong, namely that all components of the multi-band Berry connection for the projected space are everywhere zero \cite{marzari_maximally_1997}. This is vastly stronger than the criterion for exponential localisation, being only that the Chern number (the \textit{BZ average band trace} of the Berry connection) vanishes.}. 

Furthermore, unlike the previous assumptions where we could easily include more terms or intermediate states, there is no similar way to remedy the error from the TBA, as adding even weak off-diagonal terms immediately ruins the nice properties of the assumption. Hence, finding ways to improve the error from the TBA is of great interest. While there is no hope of diagonalising the position operator exactly, even for large projected spaces, we can chose a basis minimizing the square of off-diagonal elements. It turns out that this is precisely the MLWFs, as is well-known and discussed in App. \ref{app:MLWFs}. Hence, using these for constructing tight-binding models is well motivated, and they are easily obtained from \texttt{Wannier90} \cite{mostofi_updated_2014}. In the next sections, we will explore ways to mitigate the error introduced by the TBA without recomputing the hoppings based on the MLWFs.

\section{Theory\label{sec:theory}}
\subsection{The tight-binding error}

The Wannier functions allow us to quantify the error made with the TBA. In \eqref{eq:HamA} we saw that the Berry connection can be decomposed in terms of a covariant term $\Um^\dagger \tilde A_j \Um$ plus a correction $\Um^\dagger\ptl_j \Um$ depending only on the geometry-independent transformation $\Um$ diagonalising the KPC Hamiltonian. 
The error $\mathbbm{a}$ in the Berry connection caused by making the TBA, sometimes called the external Berry connection \cite{ibanez-azpiroz_ab_2018}, is then 
\begin{equation} \label{eq:OD_error_H}
    \mathbbm{a}_j = \Um^\dagger \tilde{\mathbbm{a}}_j \Um
\end{equation}
where
\begin{equation} \label{eq:OD_error_W}
    \tilde{\mathbbm{a}}_j^{\a\b}(\tau_j^\a) = \sum_\R e^{i\kv\cdot\R}\braket{\a\Ov | \hat r_j -\tau^\a_j | \b\R}.
\end{equation}
We can consider how this error changes with positions $\tau^\a_j$, as suggested by the functional notation above. This reflects the fact that, when choosing a diagonal tight-binding model, we must make a choice for what diagonal matrix to pull out of $\braket{\a \Ov| \hat \rv| \b\R}$.

A natural choice for positions is to minimize the squared Frobenius norm of \eqref{eq:OD_error_H} summed over the BZ. It is straightforward to see that we can at most remove the home cell diagonal contributions so that the WCCs are optimal under this norm (a formal minimization is given in App. \ref{app:MLWFs}). Hence, the MLWF WCCs are optimal for minimizing the TBA error in the Berry connection for the full projected space. As such, they have been the standard choice in prior literature \cite{ibanez-azpiroz_ab_2018}. 

However, these positions, minimizing the error in the position operator and the projected-space Berry connection, are \textit{not} the same as those minimizing the error in physical observables. Indeed, there are physical reasons why certain terms are more significant than others. For example, observables in insulators, like optical responses, typically depend mainly on the rectangular occupied-unoccupied block of $\Am_j$, which we will call the interspace Berry connection. It is therefore sensible to use information, such as occupations or data from experiment or ab-initio computations, to tune the orbital embeddings so that our minimal model most accurately captures the physics of interest.

%%%%%%%%%%%%%%%%%%%% [This is where we start presenting new things]%%%%%%%%%%%%%%%%%%%%%%%%%%%%%%%%%%%

\subsection{Optimally embedded tight binding} 

In principle, we can find optimal embeddings for any given observable, defined as the positions $\tau_j^\a$ minimizing the mean squared error on that observable as compared to some baseline given by either experiment or first-principles computations.

As is often true of minimal models, we face a trade-off between accuracy and generality. For many applications, we have some idea of what class of observables we want our model to fit, and we can use this information in choosing the embedding. Below, we present two general and natural improvements over the WCCs for reproducing geometry-dependent observables before discussing how to fit specific optical responses.

\subsubsection{The interspace Berry connection}

For insulators, it is natural to use the occupations to inform our choice of embeddings. We do this most naturally by fitting the interspace Berry connection, which differes from the full Berry connection in that it only includes the rectangular occupied-unoccupied block. This proves highly practical, as we can find a direct analytical least squares solution for the optimal embeddings, and the needed matrices are immediately available from \texttt{Wannier90}, using the \texttt{write\_hr} and \texttt{write\_r} flags in the \texttt{.win}-file.
The exact expression for the embedding choice which best fits the interspace Berry connection is derived in App. \ref{app:interspace_berry} in terms of a pseudoinverse equation $\mathcal{H} \bm \tau= \vec{y}$.

While practical, this embedding is in our testing close to the MLWF WCCs (recall that this emdedding minimises the \textit{total} Berry connection error for the projected space, disregarding occupations) and so the improvements on responses are minor. Furthermore, the Berry connection is not a nice gauge-invariant object and relies on Wannier functions for the reference response. Using an experimental observable is preferable so that the approach can also be applied in workflows which do not rely on constructing Wannier functions.

\subsubsection{The quantum geometric tensor}

The most obvious gauge-invariant geometric quantity to fit is the Abelian QGT of some isolated subspace of interest, usually the occupied band subspace,
\[
    \qgt_{ij} = \sumNM \qgt^{nm}_{ij},
\]
here written in terms of the optical QGT
\[
    \qgt^{nm}_{ij} = \braket{\ptl_i u_n|u_m}\braket{u_m|\ptl_j u_n} = A^{nm}_i A^{mn}_j
\]
which enters optical responses. 
The Abelian QGT captures both the Fubini-Study metric and the Berry curvature in its real and imaginary parts, respectively.

It is worth noting that the Abelian QGT can be computed directly from plane-wave DFT on the coarse $\q$-grid without any finite difference approximation \cite{kim_direct_2025} for systems without spin orbit coupling (SOC), as discussed at the end of App. \ref{app:qgt}. Still, unless the DFT-grid is already somewhat dense, it is advisable to use a Wannier interpolated QGT as it often exhibits sharp features in $\kv$-space. Importantly, computing the QGT from the Wannier interpolated Berry connection works equally well for the SOC case.

In App. \ref{app:qgt} we show how to solve for the optimal embedding for both the Abelian and the optical QGT. Using the geometry decomposition presented in the next section, we compute and provide the Jacobian and Hessian, and find in our testing that the fit is very cheap. We note that the same approach can be used to quickly obtain the embeddings minimizing the metric trace, as is relevant for computing superfluid weight \cite{huhtinen_revisiting_2022}, or to study the role of geometry in tight binding more generally, as discussed in Sec.~\ref{sec:geometry}.

\subsubsection{\label{sssec:optical_responses}Optical responses}
Using the procedure for fitting the orbital embeddings to the optical QGT, with the interpolated Berry connection from \texttt{Wannier90}, we can also cheaply obtain specific fits for other common first and second-order optical responses.

As an example, the optical conductivity is given by \cite{ahn_riemannian_2022}
\begin{equation} \label{eq:linear_response}
    \varsigma_{ij}(\omega) = \frac{\pi e^2}{h}\int_\kv \sum_{m,n} f_{nm}Q^{nm}_{ij} \omega_{mn} \delta(\omega-\omega_{mn})
\end{equation}
where $-e$ is the electron charge and $h = 2\pi \hbar$ is Planck's constant, while $\omega_{mn} = (E_m - E_n)/\hbar$ and $f_{nm}$ is the Fermi occupation difference between bands $n$ and $m$. For positive frequencies at zero temperature, only the terms with $n$ occupied and $m$ unoccupied contribute, giving
\[
   \varsigma_{ij} = \frac{\pi e^2}{h}\int_\kv \sumNM
    Q_{ij}^{nm} \om_{mn} \delta(\om -\om_{mn}).
\]
The position derivatives used for the fit are hence just a weighted sum of the derivatives of the optical QGT from before. 

The injection current takes a similar form, though the geometry independent weights are dispersions rather than frequencies \cite{lihm_comprehensive_2022}:
\[
    \eta_{kij} = -\tau \frac{\pi e^3}{\hbar^2} \int_\kv \sum_{\substack{n\,\mathrm{occ} \\ 
                    m\,\mathrm{unocc}}} 
    Q_{ij}^{nm} \ptl_k\omega_{mn} \delta(\om - \om_{mn}).
\]
Here, $\tau$ is a phenomenological relaxation time. Once again, we only need the position derivatives for the optical QGT for the fit.

The shift current \cite{sipe_second-order_2000,ibanez-azpiroz_ab_2018} 
\begin{align*}
    \sigma_{kij}& = -\frac{\pi e^3}{4\hbar^2} \int_\kv \sum_{n,m} f_{nm} i(I_{kij}^{mn} + I_{kji}^{mn})\\
    &\hspace{65pt}\times\left[\delta(\omega - \omega_{mn}) +\delta(\omega - \omega_{nm})  \right] \\
    =& \:\frac{\pi e^3}{2\hbar^2} \int_\kv \
    \sum_{\substack{n\,\mathrm{occ} \\ 
                    m\,\mathrm{unocc}}} 
     \Im \left[I^{mn}_{kij} + I_{kji}^{mn} \right]\,\delta(\om - \om_{mn})
    % \sigma_{kij}& = -\frac{\pi e^3}{4\hbar^2} \int_\kv \sum_{n,m} f_{nm} iQ^{nm}_{ij} (R^{k,i}_{mn} + R^{k,j}_{nm}) \\
    % &\hspace{65pt}\times\left[\delta(\omega - \omega_{mn}) +\delta(\omega - \omega_{nm})  \right] \\
    % =& \:\frac{\pi e^3}{2\hbar^2} \int_\kv \
    % \sum_{\substack{n\,\mathrm{occ} \\ 
    %                 m\,\mathrm{unocc}}} 
    %  \Im \left[Q^{nm}_{ij} (R^{k,i}_{mn} + R^{k,j}_{nm}) \right]\:\delta(\om - \om_{mn})
\end{align*}
has additional geometry dependence, depending on
\[
    I_{kij}^{mn} = -iQ^{mn}_{ij}R_{kj}^{nm},
\]
written here in terms of the optical QGT and the (complex) shift vector
\[
    R_{kj}^{mn} = A_k^{mm} - A_k^{nn} +i\ptl_k \ln A_j^{mn} \quad (m\neq n\: \text{else }0)
\]
which is gauge invariant since the gauge correction from the last term cancels those from the first two, and anti-Hermitian in the upper indices. We refer to \cite{ibanez-azpiroz_ab_2018} for technical details regarding how to evaluate this in practice \footnote{Implementations are directly available in WannierBerri \cite{tsirkin_high_2021} or postw90 \cite{mostofi_updated_2014}.}. Analytic position derivatives for this geometric object can in principle be computed using the method presented in the next subsection to dramatically reduce computational cost, but we use numerical derivatives in practice.

\subsubsection{The optimal embedding procedure} 

We reiterate that optimal embeddings can equally well be obtained for other responses (e.g. viscoelastic effects \cite{viscoelastic}) as long as the tight-binding expression for the response can be recast in terms of standard geometric quantities for which one can compute the geometry dependence as discussed shortly.

As such, we can summarise the procedure for obtaining optimal embeddings as follows: (i) Obtain a reference observable $\bm y_0$ for a geometric quantity which can be computed in tight binding as $\hat{\bm y} (\tau^\a_j)$. Note that $\bm y_0$ may have multiple indices and depend on other parameters such as $\kv$ or $\om$. (ii) Make the geometry decomposition as described in the next section and take derivatives with respect to positions $\tau^\a_j$ to get the Jacobian and Hessian (this is described in detail for the QGT in App. \ref{app:qgt}). (iii) Finally, run a standard MSE minimisation of the error $\|\hat y - y_0 \|$ using the obtained derivatives \footnote{For the QGT, our fits typically take less than a second on a laptop for simple real materials, but do note that cost landscapes can vary a lot depending on the observable.}.

\subsection{Geometry decomposition\label{ssec:decomposition}}

We show a simple and general approach for decomposing tight-binding expressions into a geometry dependent and independent part. 
This lets us minimize tight-binding quantities with respect to orbital embeddings, which is directly physically relevant since computing the minimal metric trace is necessary to get accurate predictions for the superfluid weight \cite{huhtinen_revisiting_2022}. Since we can take first and second position derivatives for arbitrary tight-binding observables, minimization is very cheap.

In extension of this, if we also have some reference value for an observable from experiment or ab-initio computation, the geometry decomposition lets us fit optimally embedded tight-binding models against this observable. In some cases, such as when there are few orbitals or strong symmetry constrains on the position degrees of freedom (DOFs), we can also solve directly for the optimal (or minimal) embeddings. 

Notice that the geometry decomposition gives the precise analytical expression constraining a tight-binding observable for fixed hoppings: If the difference between ab-initio results and the geometry independent term cannot be approximately represented by this expression, we can never hope to get a good fit for the observable with the given hoppings. Since the position DOFs scale only linearly with the number of orbitals and are entirely $\kv$-independent, this is quite restrictive. For $\kv$-dependent observables, the optimal embeddings are overdetermined even by a coarse grid, and if no good fit is possible on that grid, we can conclude that the TBA should not be applied on that model.

The geometry decomposition also clarifies where geometric effects are expected to be important, and will be central in our isolated study of geometric effects in Sec.~\ref{sec:geometry}.

\subsubsection{Decomposing the Berry connection}

Consider an orbital displacement $\d \bm \tau_\a = \bm \tau_\a -\bm \tau'_\a$ from primed to unprimed positions. The diagonalising matrix of the TBC Hamiltonian, in \eqref{eq:V}, then transforms as
\[
    V_{\a n} =  e^{-i\kv\cdot\d\bm\tau_\a} V'_{\a n}.
\]
It is convenient to introduce diagonal displacement matrices $\D_j^{\a\b} = \d \tau^\a_j \delta_{\a\b}$ and the unitary matrix 
\[
\Dm =  e^{-i\kv \cdot \bm \D} \quad \text{such that} 
\quad\ket{u_n} = \Dm\ket{u'_n}.
%\quad \vv_n = \Dm \vv_n'.
\]
(Here, the displacement operator strictly acts by matrix multiplication on the coefficients $\vv_n = \Dm \vv_n'$, which play the role of the states $\ket{u_n}$ in tight-binding expressions in the TBC as discussed in Sec.~\ref{sssec:TBA}. We will stick with standard ket-notation, as is conventional. Do note that the displacement operator is neither a gauge nor basis choice but, as we shall see, alters the physical system and observables.)

The chain rule gives
\[
    \ket{\ptl_j u_n} = \ptl_j \Dm \ket{u'_n} + \Dm \ket{\ptl_j u'_n} = \Dm (-i\D_j\ket{u'_n} + \ket{\ptl_j u'_n}), 
\]
so the Berry connection transforms as
\begin{align*}
    A_j^{\,nm} &\equiv \braket{u_n|i\ptl_j u_m} \\
    &= \bra{u'_n}\Dm^\dagger i \Dm\lb-i\D_j\ket{u'_m}+\ket{\ptl_j u'_m}\rb \\
    &= A'^{\,nm}_j +\D_j^{nm},
\end{align*}
where we introduced the notation $\D_j^{nm} = \braket{u'_n|\D_j|u'_m}$. This gives the multiband generalisation of Eq.~(12) in \cite{simon_contrasting_2020}. 

In the special case that the original (primed) positions were all zero, we get $\d \tau^\a_j = \tau_j^\a$ and hence a decomposition of any tight-binding quantity 
\begin{equation} \label{eq:decomposition}
A_i(\tau_j^\a) = A'_i
+ \d A_i(\tau_j^\a),
\end{equation}
where the prime now marks the geometry independent term and the $\d$-term is the geometry dependent correction in the KPC.

\subsubsection{Decomposing the quantum geometric tensor}
Applying this to the optical QGT, we get
\begin{align} \label{eq:dQ_opt}
   \d \qgt_{ij}^{nm} &\equiv A^{nm}_i A^{mn}_j - A'^{\,nm}_i A'^{\,mn}_j \\
    &= \D_i^{nm} A'^{\,mn}_j +A'^{\,nm}_i \D_j^{mn} +  \D_i^{nm} \D_j^{mn} \notag.
\end{align}
Summing over $n$ occupied and $m$ unoccupied, we get the displacement for the Abelian QGT for the occupied subspace. (More generally, we can let $n$ run over any isolated group of interest and $m$ run over intermediate states.)

The case of a single occupied band takes a particularly simple form, since the sum over intermediate states reduces to the complimentary projection $1-\ket{u_n}\bra{u_n}$. After some simple algebra, we find
\begin{align}
    \d\qgt_{ij}^n &= \sum_{m\neq n}  \d \qgt^{nm}_{ij} \label{eq:1band_qgt} \\
    %&= \sum_{m\neq n} \lb \D^{nm}_i A^{mn}_j +A^{nm}_i \D^{mn}_j +  \D^{nm}_i \D^{mn}_j \rb \\
    % &= i\braket{\uu{n}|\D_i(1-P_n)|\ptl_j\uu{n}} -i\braket{\ptl_i\uu{n}|(1-P_n)\D_j|\uu{n}}  + \braket{\uu{n}|\D_i(1-P_n)\D_j|\uu{n}}\\
    &= i \left[ \braket{u'_n|\overline{\D_i^n}|\ptl_j u'_n} - \braket{\ptl_i u'_n|\overline{\D_j^n}|u'_n} \right] +  C^n_{ij}, \notag
\end{align}
where the centred displacement $\overline{\D_j^n} = \D_i - \D_i^{nn}$ and the covariance
\[
    C^n_{ij} = \braket{u'_n|\D_i \D_j | u'_n} - \D_i^{nn}\D_j^{nn}
\]
were introduced. Written in this way, invariance under uniform shifts, corresponding to a different choice of origin, is clear \footnote{This shift-invariance holds also for the multi-band case, as follows from \eqref{eq:dQ_opt} noting that $\D^{nm}_{j} = c_j \d^{nm}$ for $c_j$ a constant is zero always when summing over complimentary subspaces $n \neq m$.}. 

We can take real and imaginary parts to obtain the geometry dependence of the Fubini-Study metric and the Berry curvature. From the latter, we can e.g. reproduce the result from \cite{simon_contrasting_2020} that the Chern number is geometry independent, since
\begin{align*}
    \d \Om_{ij}^n &= -2\,\Im \d \qgt_{ij}^n \\
    &= -2\,\Re \left[ \braket{u'_n|\D_i|\ptl_j u'_n} - \braket{\ptl_i u'_n|\D_j|u'_n} \right] \\
    &=\ptl_i\D^{nn}_j -\ptl_j \D^{nn}_i,
\end{align*}
where we in the second line used that the centring terms $\D_i^{nn}A'^{\,nn}_j$ and covariance are purely real. To get the last line, we expanded the real part as a sum of complex conjugates. Integrating this exact form over the boundary-less BZ gives zero by Stokes' theorem. The multi-band generalisation is similar and given in App. \ref{app:Chern}.

In fact, geometry independence of the Chern number extends beyond the TBA, since the Chern number is a topological invariant of the bundle associated with an isolated group of eigenspaces of the Hamiltonian. The position operator only affects the choice of connection on this bundle and makes no difference to its topology.

\section{Fitting optical responses\label{sec:optimal}} % appl. I

%%%%%%%%%%%%%%%%%%%%%%%%%%%%%%%%%
%%%%%%%%%%%%%%%% SHIFT CURRENT
%%%%%%%%%%%%%%%%%%%%%%%%%%%%%%%%%
\begin{figure*}[!t]
    \centering
    \def\svgwidth{\linewidth}
    \input{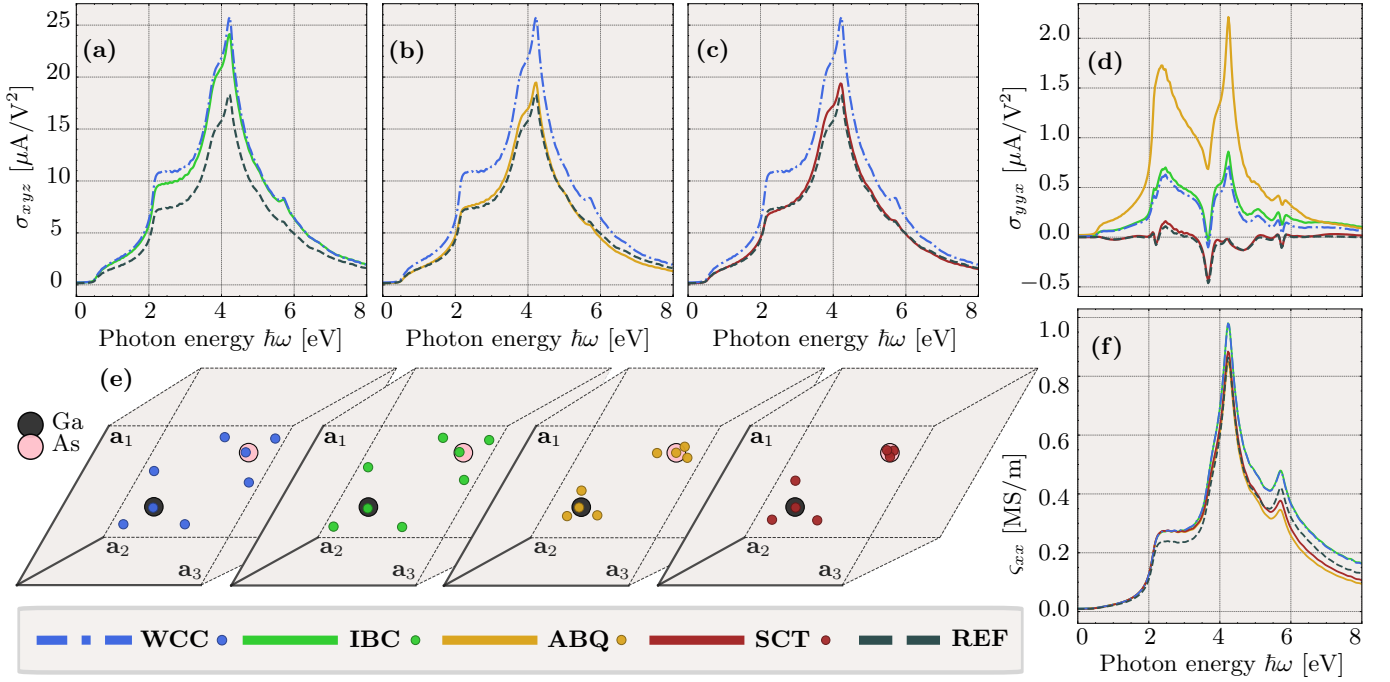}
    \caption{GaAs tight biding optical responses for different embeddings. 
    Legend abbreviations denote: Wannier Charge Centers (WCC), Interspace Berry Connection (IBC), Abelian QGT (ABQ), Shift Current Tensor (SCT) and Wannier interpolation (REF).  
    \textbf{(a)}--\textbf{(c)} The only independent shift current component ($xyz$). 
    \textbf{(d)} A symmetry-disallowed shift current component ($yyx$). 
    \textbf{(e)} Orbital embeddings in the irreducible unit cell. 
    \textbf{(f)} The only independent optical conductivity component ($xx$).
    }
    \label{fig:GaAsSC}
\end{figure*}

Having derived how tight-binding expressions for optical responses depend on geometry, we now apply this theory to study optimal embeddings for optical responses for 
GaAs (zincblende) and CdS (wurtzite). GaAs was one of the first materials whose shift current spectrum was obtained via band-structure computation \cite{sipe_second-order_2000}. Previous studies of GaAs shift current, without optimal embeddings, found poor quantitative agreement between the TBA and Wannier interpolation \cite{ibanez-azpiroz_ab_2018}. CdS (wurtzite) is one of the simplest materials with a symmetry-allowed injection current component \cite{nastos_optical_2010}.

Note that the accuracy of the ab-initio computations is of only secondary importance for our purposes, as we are primarily interested in comparing differently embedded tight-binding models against DFT taken as ground truth. 
As such, we are justified in making minor approximations which we do not expect to qualitatively change results. A thorough description of computational details can be found in App. \ref{app:computation}, including on DFT and Wannier interpolation as well as numerical details on obtaining and fitting optical responses.

\subsection{Results\label{ssec:optical_results}}
\subsubsection{Optical responses for GaAs}

Previous literature has found that the TBA gives a large quantitative error on shift current for GaAs, using the MLWF WCC embedding \cite{ibanez-azpiroz_ab_2018}. In Fig.~\ref{fig:GaAsSC}, we show that we can tune the orbital embeddings to make this error negligible.
    
Panels (a)--(c) show the only independent GaAs shift current component $\sigma_{xyz}$ for the MLWF WCCs (WCC) and reference response (REF, obtained from full Wannier interpolation) versus responses obtained by fitting the orbital embeddings to optimally reproduce (a) the interspace Berry connection (IBC), (b) the Abelian QGT (ABQ) and (c) all components of the shift current tensor (SCT). The IBC fit provides only a minor correction, while the ABQ and SCT fits are excellent. The tailored SCT fit is perhaps ever so slightly better at the plateau around $3\,$eV and at high frequencies. 

More noticeably, the tailored embeddings do better than the others in suppressing the symmetry-disallowed components, as seen in panel (d) where the symmetry-disallowed $yyx$-component is shown. Notice that the y-axis differs by an order of magnitude here, so that all embeddings significantly suppress this component. Even the Wannier interpolation and SCT-response do not give exactly zero, however, likely due to slight symmetry-breaking in the disentanglement procedure. 

The small symmetry-disallowed components contribute very little to the optimization cost, so it might be surprising that the optimal embeddings give such a good fit to these components also. This indicates that excess DOFs in the orbital embeddings remain even after fitting the independent non-zero shift current component.

The ABQ-response is an order of magnitude larger than SCT on the symmetry-disallowed component. The symmetry breaking is seen clearly in panel (e), where we plot the embeddings in the irreducible unit cell. This highlights that symmetry preservation is not guaranteed in numerical optimisation. Still, the other embeddings seem to retain symmetry, indicating that restricting the optimisation region to the symmetry allowed region might give faster yet equally accurate results. For most applications, the additional effort of finding the symmetry projectors will be excessive as the direct fit is generally good, but symmetry adapted fits remains an interesting suggestion for making minimal analytical models going forward.

%%%%%%%%%%%%%%%%%%%%%%%%%%%%%%%%%
%%%%%%%%%%%%%%%% OPTICAL CONDUCTIVITY
%%%%%%%%%%%%%%%%%%%%%%%%%%%%%%%%%

Finally, panel (f) shows the only independent symmetry-allowed (first-order) optical conductivity component $\varsigma_{xx}$. While the WCC-response is already decent here, the embeddings which fit the shift current also generalise to give a substantial improvement on this response.

\begin{figure}[h]
    \centering
    \def\svgwidth{1\linewidth}
    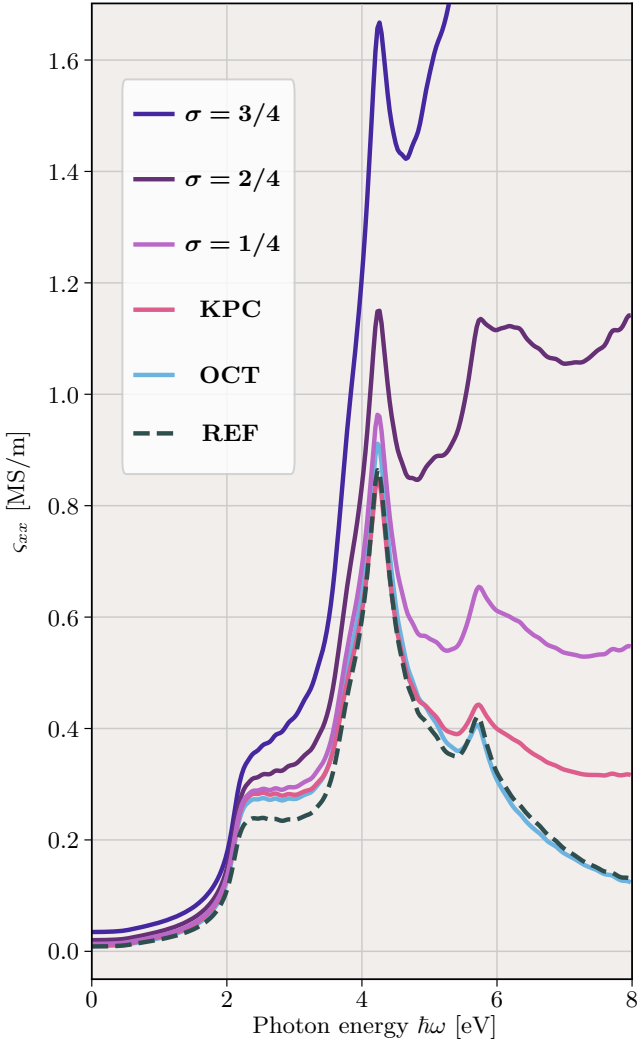
    \caption{
    Optical conductivity $\varsigma_{xx}$ for GaAs for different embeddings. Beyond the tailored fit on the Optical Conductivity Tensor (OCT) and Wannier interpolation (REF), we show plots for increasing noise, corresponding to the average of 15 responses computed from orbital embeddings drawn from a normal distribution with standard deviation $\sigma$, and the geometry independent contribution (KPC) obtained by placing all orbitals at the origin (i.e. $\sigma = 0$).}
    \label{fig:naive_random_oc}
\end{figure}

In Fig.~\ref{fig:naive_random_oc} we show the optical conductivity of GaAs as predicted from positions corresponding to a tailored fit for the optical conductivity tensor, as well as random positions with increasing levels of noise. There are multiple interesting things to note. First of all, the low frequency behaviour is only weakly geometry independent in tight binding, with the plateau around $2\,$eV stable even at noise with standard deviation of about a quarter of a unit cell diameter. At the same time, the low-frequency off-diagonal error is not negligible, so that the fit cannot be made perfect even with optimal embeddings. 

The frequencies at which the response peaks are also extremely robust, though the heights fluctuate. Furthermore, the plot illustrates the dramatic errors caused by a bad choice of embedding, and the importance of the geometry dependent correction terms in the KPC \footnote{The behaviour with noise here is expected, since it is know from \eqref{eq:opt_bound} that the area under the optical response is related to the quantum weight, which blows up with the metric trace as the orbitals move apart.}.

%%%%%%%%%%%%%%%%%%%%%%%%
%%%%%%%%%%%%% CdS
%%%%%%%%%%%%%%%%%%%%%%%%%

\subsubsection{Optical responses for CdS}

\begin{figure*}[!t]
    \centering
    \def\svgwidth{1\linewidth}
    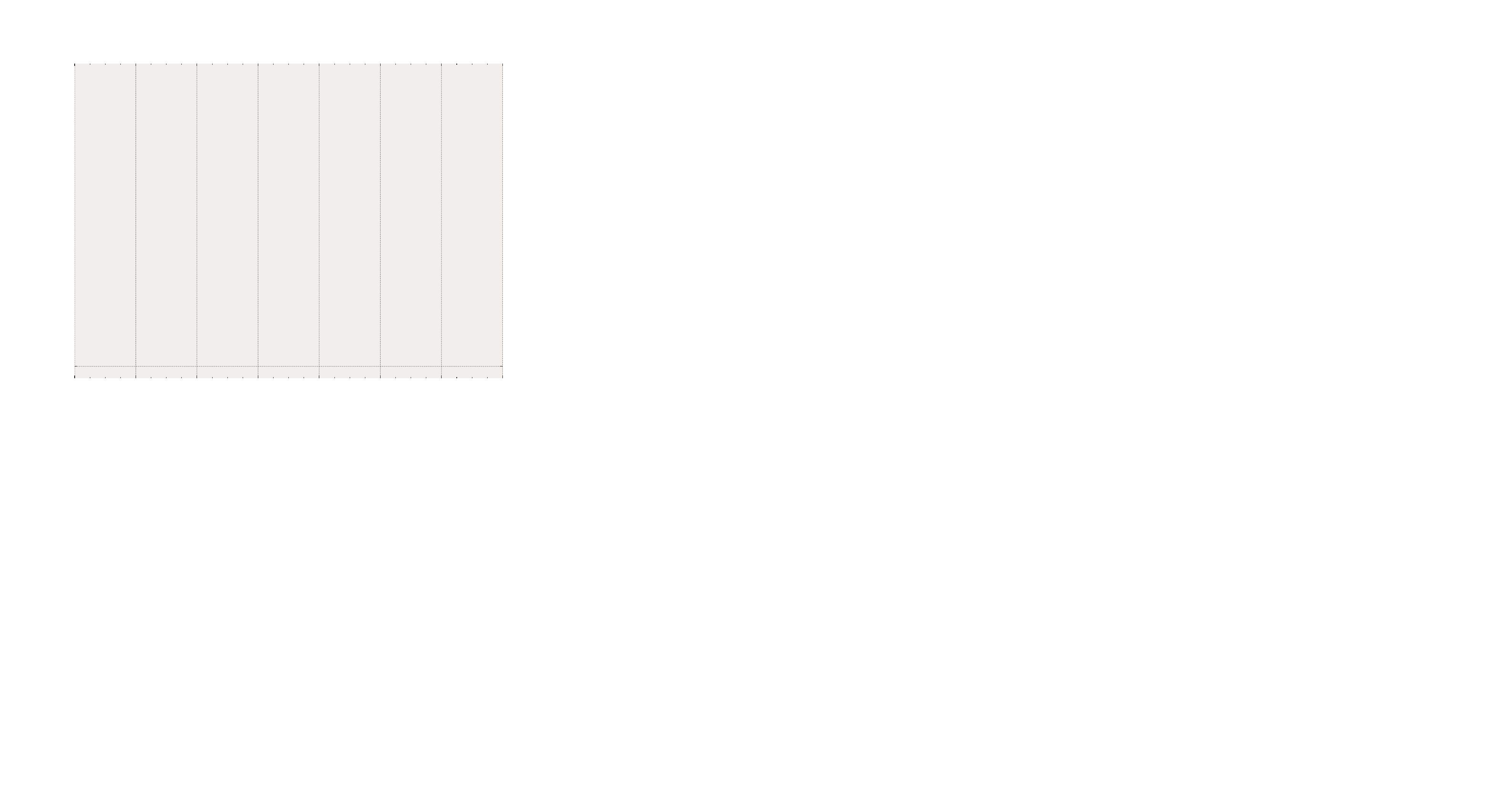
    \caption{CdS tight-binding optical responses for different embeddings. Legend abbreviations denote: Wannier Charge Centers (WCC), Interspace Berry Connection (IBC), Optical Conducitvity Tensor (OCT), Injection Current Tensor (ICT), Shift Current $xxz$-Component (SCC) and Wannier interpolation (REF). 
    \textbf{(a)}--\textbf{(b)} The independent optical conductivity components (resp. $xx$ and $zz$). 
    \textbf{(c)} The independent injection current component ($xzx$). The scaling is arbitrary, with the relaxation time set to $1\,$fs.
    \textbf{(d)} One of three independent shift current components ($xxz$).
    \textbf{(e)}--\textbf{(f)} The wurtzite structure, with Cd in grey and S in yellow. In (e), a top view of a single layer above the $xy$-plane. The lattice extends vertically for each irreducible unit cell as illustrated in (f) for the cell above the red dotted line. 
    \textbf{(g)} The tight-binding model (TBM) band structure, agreeing perfectly with DFT below the energy cut-off $E_m$.}
    \label{fig:CdS_response}
\end{figure*}

The results for CdS (wurtzite structure) are shown in Fig.~\ref{fig:CdS_response}. In panels (a)--(b) we show the independent components $\varsigma_{xx}$ and $\varsigma_{zz}$ of the optical conductivity. Note that MLWF WCCs display a reasonable quantitative error, counter to expectations expressed in prior literature that first order responses are accurately captured in the TBA \cite{wang_first-principles_2017}. The optimal embedding does better, though we find again the low-frequency error is only weakly geometry dependent, and therefore cannot be corrected.

The independent component of the injection current, $\eta_{xzx}$, is shown in panel (c). Here, the IBC fit captures the peak behaviour better than the MLWF WCCs, but does slightly worse at low frequencies. The tailored ICT fit is excellent.

CdS (wurtzite) has three independent shift current components $\sigma_{ijk}$ ($xxz$, $zxx$ and $zzz$). We find that the MLWF WCCs do well at low frequencies for all responses, and give a qualitatively correct fit for the $zxx$- and $zzz$-components. However, we find a notable qualitative error from around $6\,$eV for the $xxz$-component, shown in panel (d). This goes beyond the quantitative errors reported in previous literature \cite{ibanez-azpiroz_ab_2018}, and serves to strengthen the point that significant care should be taken in considering embeddings when predicting shift currents with the TBA. While we can find an optimal embedding fitting the other two components simultaneously, it seems we cannot qualitatively change the shape of $\sigma_{xxz}$ without simultaneously altering the shapes for the other, qualitatively correct, responses. Hence, while a reasonable component-wise fit is possible, labelled SCC in panel (d), we find that we cannot fit all shift current components simultaneously. This might indicate that a qualitative error with the MLWF WCCs on one observable (here, $\sigma_{xxz}$) can imply that the TBA, even with optimal embeddings, might generalise poorly beyond that observable. In such a case, then, using the full position operator may be advisable.

\subsection{Discussion}

\subsubsection{Symmetry considerations\label{sssec:symmetry}}

In Sec.~\ref{sssec:approximations}, we noted that disentanglement and the TBA can break symmetries. Additionally, numerical optimisation can generally cause minor symmetry breaking, such as maximal localisation slightly opening Kramers-degenerate band touchings in systems with time-reversal symmetry. When symmetries are of particular importance, such as for analytical minimal models, where we want to utilise as much physical information about the system as possible, symmetry adapted Wannier functions may be used, forcing the WCCs to high-symmetry positions.

Frequently, there will be sufficiently many orbitals in a model that free Wyckoff parameters remain even in the symmetrised orbital embeddings. Since symmetry adapted Wannier functions may be less localised than freely MLWFs, and hence give larger errors with the TBA, it may be especially wise in this case to tune these Wyckoff DOFs to mitigate geometric errors in diagonal tight binding.

A symmetry adapted optimal embedding gives a physically motivated way to fix these free parameters, by fitting an observable of interest as discussed above. The symmetry adapted hoppings serve to label the orbitals into Wyckoff groups, and so their positions $\tau^\a_j(x_\mu)$ can be written in terms of the Wyckoff parameters $x_\mu$, obtained e.g. from the Bilbao Crystallographic Server \cite{aroyo_bilbao_2006}. The optimal embeddings for any geometry dependent response $f(\tau^\a_j)$ can then be obtained with the procedure presented in this work, with derivatives of $f(\tau^\a_j)$ obtained in the same way as for the QGT in App.~\ref{app:qgt} and then using the chain rule to obtain derivatives of $f(\tau^\a_j(x_\mu))$ with respect to the Wyckoff parameters $x_\mu$. While symmetries reduce the DOFs for the optimal embedding dramatically, we might hope that the optimal embedding will respect symmetries (at least those relevant to the observable in question). Hence, the reduction in complexity might not come at a large cost in flexibility, but rather serve positively to reduce the optimisation cost further.

Small or highly symmetric systems, may have no free Wyckoff parameters, implying that there exists a unique choice for symmetrized optimal embeddings. Other materials however, such as wurtzites, will typically have free parameters. For small analytical models, symmetry constraints might make it feasible to solve for the optimal embedding analytically based on one or two experimental parameters (such as e.g. the measured capacitance \cite{komissarov_quantum_2024}).

\subsubsection{Physical interpretation}

When viewing diagonal tight binding as an approximation of a more complete Wannier picture, the optimal embeddings can be interpreted simply as a free parameter in the minimal model which should be fixed to minimize the error in the approximation. The orbitals $\ket{\a\R}$ are still considered to correspond to Wannier functions, and the TBA amounts to taking an effective position operator, first by neglecting off-diagonal terms in the position matrix and second by perturbing the diagonal terms. The surprising success of tight binding is then partially explained by exponential localisation. This conservative view is solid and sufficient to state and appreciate all the results presented in this paper.

However, diagonal tight binding is often considered an excellent model of simple condensed matter systems in its own right, and the model parameters admit natural interpretations beyond being approximations to the Wannier picture. This is particularly the case in theoretical work which is often based on a tight-binding description. If one is given an effective hopping Hamiltonian and makes a length-gauge perturbation, as is relevant to optical responses, the orbital embeddings can be naturally interpreted as a set of Dirac-delta-like position eigenstates $\ket{\R + \bm\tau_\a} \equiv \ket{\a \R}$ carrying the interpretation of possible collapse sites of electrons when measured, e.g. by a photon. A good choice of orbital embeddings then corresponds to the choice of positions which most faithfully represent the electron-photon interaction. This view hence highlights the vital importance of choosing the possible electron sites with care if one wishes to accurately reproduce optical responses.

With such a physical interpretation of the diagonal tight-binding model, the MLWFs -- which are in any case just one particular basis choice obtained from a (natural yet not inherently physical) mathematical optimisation designed in part to minimize long-range hoppings -- serve us only as one among several possible ways of obtaining an effective Hamiltonian which adequately reproduces the band structure (other options are e.g. Slater-Koster, LCAO or machine learning methods). The tight-binding error remains well-defined as compared to any gauge-invariant geometric observable from e.g. experiment, like the QGT \cite{kim_direct_2025} or optical responses, and hence optimal embeddings remain equally applicable for tight-binding models independent of whether they were obtained through real-space representations of Wannier functions or by entirely different methods. 

Note that this physical position eigenstate interpretation corresponds to keeping wave-function values, corresponding to the expansion coefficients of each orbital site, only at a sparse discrete set of points. This underscores the dramatic simplification made in tight binding as compared to the continuum picture. Finally, the interpretation makes the slightly peculiar role of the displacement operator from Sec.~\ref{ssec:decomposition} more transparent: It implies that the displacement operator changes the possible sites to which electrons are allowed to collapse, and hence affects the electron-photon interaction directly. This is clearly neither a gauge nor a basis choice but a physical perturbation to the system. Still, while such an operator might be conceivable, it is then also partially a modelling artifact. We discuss the challenges of realising a displacement operator in an experimental setting, as well as why one might want to do so, further in Sec.~\ref{ssec:geo_discussion}.

\section{Qualitative effects of geometry\label{sec:geometry}}

As a second application of geometry decomposition and position derivatives, we study qualitative effects of geometry in tight-binding more generally. First, we make some theoretical predictions on where geometric effects are expected to be important and discuss the geometry dependence of toy models. We then present results for the prototypical Chern insulator V$_2$O$_3$ \cite{mellaerts_two_2021}. Computational details, including DFT and Wannierisation, toy model parameters and numerical details, can be found in App. \ref{app:computation}.

\subsection{Theory}

%%%%%%%%%%%%
%%%%%%%%%%%%%%%%%%
%%%%%%%%%%%%%%%%%%%%%%%%%%%%%%

Artificially freezing the hoppings lets us investigate theoretically the isolated effect of the position operator for systems adequately modelled by tight binding. 
The main motivation for doing this is to understand where geometric effects are important and where they are less so. In practice, electronic structure computations are typically validated by comparing with band structures, which might be sufficient when the geometric effects on the quantity of interest are small. However, if we know that geometric effects can be large, we might want to use a geometric quantity like an optical response for additional validation. In such cases, full Wannier interpolation is a principled approach, but when tight binding is preferred, optimal embeddings are attractive to improve the model.

\subsubsection{Band gap dependence\label{sssec:band_gap_dep}}
Beyond the binary subdivision of geometry dependent and independent quantities, one may consider the degree of geometry dependence. As an example, our results in Sec.~\ref{ssec:optical_results} indicate that the low frequency optical conductivity is only weakly geometry dependent. %The band gap dependence might give a partial explanation of this. 
By considering how the band gap enters geometric correction terms, we can say something about what systems are expected to have strong geometric effects and where in $\kv$-space the geometry dependence is expected to be important

For concreteness, we consider the QGT, with geometry decomposition given in \eqref{eq:dQ_opt}. By Kubo-expansion, the geometry independent contribution to the QGT varies roughly like the squared reciprocal of the band gap, $\om_g(\kv)^{-2}$. Since it involves two $\kv$-derivatives, we furthermore expect it to vary strongly with $\kv$. % would be nice to make this more rigorous
From \eqref{eq:dQ_opt} or \eqref{eq:1band_qgt}, we see that the geometry dependence of the QGT contains a cross-term involving first derivatives (proportional to the Berry-curvature) and a flat term with no $\kv$-derivatives and no direct band gap dependence. The cross-term goes like $\om_g^{-1}$ and is expected to depend less strongly on $\kv$. The flat term is expected to have the weakest $\kv$ dependence.

Generally, then, we expect the geometry independent term to dominate at small band gaps in $\kv$-space. In particular, geometry corrections become (relatively) less important for systems close to a critical point (band touching). This might give a partial explanation for the weak geometry dependence of the low frequency optical conductivity. The relevant transitions at low frequencies pick out the metric only where the band gap is smallest, which is where geometry independent terms are expected to dominate.

At moderate band gaps, we expect the geometric correction to be important locally. For a moderate or large global band gap, we expect the flat term to be important for integrated quantities like the quantum weight. 

In particular, for qualitative local effects, such as a relatively large peak in the metric trace without a relatively narrow band gap, the cross-term will be the main contributor, and we might as a first guess look for such effects in toy models or materials with a flat plateau around the minimum of the band gap. This gives a large region in the BZ where geometric effects will be relatively prominent. 

Indeed, we find for toy models like Haldane and QWZ that the geometric contributions to the local metric trace behaviour is relatively strongest when the model parameters are tuned so that the bands are as flat as possible in a region around the peak, such as in Figs.~\ref{fig:Haldane} and \ref{fig:QWZ}.

One intuitive way to understand this is that, because the dispersion $E(\boldsymbol{k})$ only varies weakly with $\boldsymbol{k}$, any changes of responses with $\boldsymbol{k}$ must be due to changes in the geometry. This is a well-known effect in e.g. flat-band superconductors \cite{peotta_superfluidity_2015,huhtinen_revisiting_2022} and also plays a key role in realizing fractional Chern insulators \cite{FCI_Flatness, FCI_LL_Geom} and other correlated phenomena arising from flat-band physics. We note that this indicates that orbital embeddings likely play a role in correlated flat-band physics starting from a tight-binding model more generally \cite{jackson_geometric_2015, lee_embedding_2025}.

\subsubsection{Bounds}

The discussion above indicates that materials with relatively flat band gaps might have a surprisingly large metric peak, compared to what one would expect from the geometry independent contribution. To test the extent of this effect, we are interested in making the metric trace large locally in $\kv$-space without altering the hoppings. 

However, for such a maximization to make sense, we must bound the optimization region, since the metric can become arbitrarily large as we push the positions apart to infinity. Luckily, there are natural (geometry independent) bounds to the quantum weight (that is, the BZ average metric trace) 
for real materials which constrain the optimisation region to a compact set \cite{onishi_quantum_2025, Experimental_QGT_bounds}. Some of these bounds are discussed in App. \ref{app:bounds}. For the applications here, though, we shall only rely on the \textit{existence} of upper bounds for the quantum weight, which motivates using equipotential surfaces of this quantity as a natural bounding constraint for the optimisation.

\subsection{Results}

From the above discussion, we are motivated to estimate the effect of orbital embeddings on the qualitative shape of the metric trace by considering how its maximum changes when the positions are varied on or inside an equipotential surface of the quantum weight. This way, the mean metric trace is held fixed, so that the orbital embeddings only \textit{redistribute} the metric trace in $\kv$-space. We will also consider limiting the orbital displacement to small perturbations from the MLWF WCC positions. Numerical details can be found in App. \ref{subapp:num_geo}.

\begin{figure}
    \centering
    \def\svgwidth{1\linewidth}
    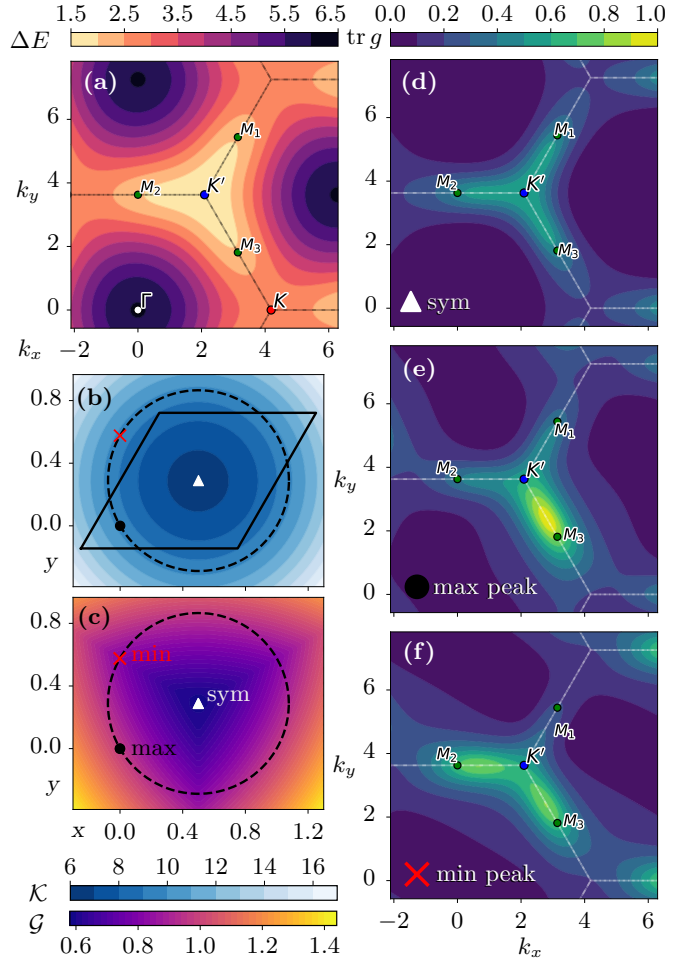
    
    \caption{Geometry dependence of the metric trace in a Haldane model with parameters $\Delta=0.3, \:t_1 = -1$ and $t_2=-0.21$, and lattice constant $a=1$. \textbf{(a)} The band gap (in units $|t_1|$) over the BZ around a $K'$ point with high symmetry points indicated. Notice the near-flat triangular plateau in the band gap between $K'$ and the C3 equivalent $M$-points marked $M_1$, $M_2$ and $M_3$. \textbf{(b)}--\textbf{(c)} The BZ average metric trace $\mathcal{K}$ as a function of orbital separation $(x,y)$ (no units), and the maximum of the metric trace $\mathcal{G}$ (units $a^2$) as a function of orbital separation, respectively. The unit cell is indicated, together with the circular equipotential curve of the quantum weight $\mathcal{K}$, corresponding to the geometry independent average metric trace contribution. Three points are indicated, corresponding to plots (d)--(f) showing the metric trace (units $a^2$) in $\kv$-space: \textbf{(d)} The white triangle marks the maximal symmetry position, \textbf{(e)} the black dot marks the origin, which is one of three maximal peak positions, and \textbf{(f)} the red cross marks one of three minimal peak position on the equipotential curve, corresponding to an equal distribution of metric trace over two $M$-points.}
    \label{fig:Haldane}
\end{figure}

\begin{figure*}[t]
    \centering
    \def\svgwidth{\linewidth}
    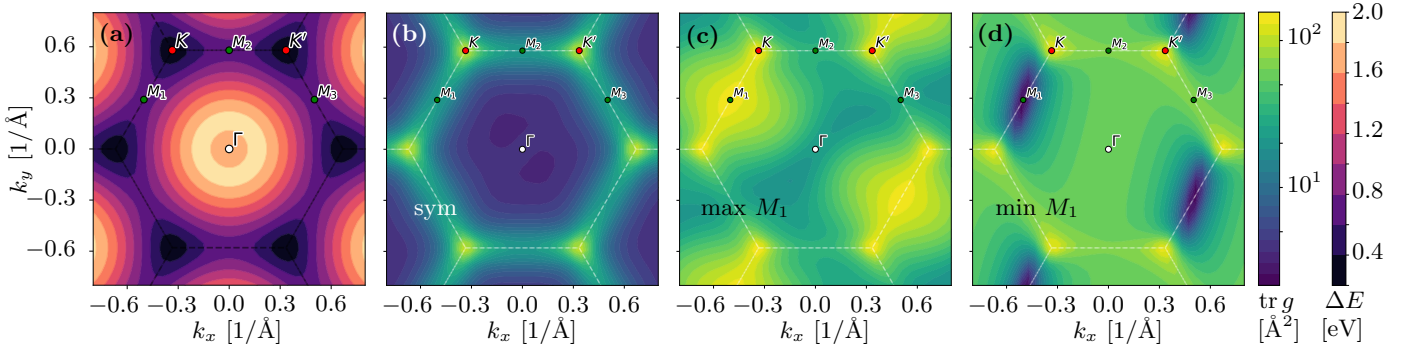
    \caption{\textbf{(a)} Band gap for monolayer V$_2$O$_3$ over the BZ with high symmetry points indicated. \textbf{(b)} Metric trace for MLWF WCCs. \textbf{(c)} Maximisation of metric trace at $M_1$ for positions constrained so that the quantum weight does not exceed its bound. \textbf{(d)} Minimization of metric trace at $M_1$ for positions constrained so that the quantum weight is exactly equal to its value in the previous panel.}
    \label{fig:V2O3}
\end{figure*}

\subsubsection{Geometry dependence in toy models\label{sssec:toy_results}}
% see theory sec

The geometry dependence of the metric trace in a Haldane model \cite{Haldane_model, jackson_geometric_2015} is illustrated in Fig.~\ref{fig:Haldane}. In panel (a), we show the band gap over the BZ. Panels (b)--(c) show, respectively, the BZ average metric trace (the quantum weight) and the maximal value of the metric trace over the BZ, corresponding to its highest peak. These are shown as a function of the orbital separation vector $(x,y)$, i.e. we fix one orbital at the origin of the unit cell and move the other orbital relative to it, without changing the hoppings. Notice that the quantum weight shown in (b) increases purely with radial displacement from the maximal symmetry position (indicated by a white triangle). The circle centered at the maximal symmetry position and running through the origin hence defines an equipotential curve for the quantum weight, as shown in panel (b). When the orbital separation vector moves along this circle, the weight of the metric trace shifts in turn towards each of the C3 equivalent $M$-points, as allowed by the position sector symmetry breaking, causing a change in the peak metric trace. In addition to the unit cell and the equipotential curve, we have highlighted three points in the real space panels corresponding to the three heatmaps (d)--(f) showing the metric trace over the BZ: The white triangle marks the C3 Wyckoff position, with metric trace in panel (d) seen to be equally distributed around each of the equivalent M-points. The black dot marks the origin (i.e. placing both orbitals on top of each other), which is the point defining the equipotential curve of the quantum weight. This is also one of three maximal peak positions, where the metric trace is pushed fully towards the $M_3$-point, as seen in panel (e). Finally, the red cross marks a minimal peak position, corresponding to the peak being equally distributed between two M-points, seen in panel (f). Similar plots for a QWZ model and a asymmetric QWZ variant are given in Figs.~\ref{fig:QWZ} and \ref{fig:QWZ_assym}.

\subsubsection{Geometry dependence in V$_2$O$_3$}

The results for maximising the metric trace, within the bounds defined by an equipotential curve of the quantum weight, at the $M_1$-point for monolayer V$_2$O$_3$ are shown in Fig.~\ref{fig:V2O3}. The first panel shows the band gap over the BZ, and the second shows the MLWF WCC metric trace. Notice the slight symmetry breaking in the WCC metric trace, underscoring that geometry dependent quantities often are more sensitive to symmetry breaking (e.g. from the Wannierisation) than the band structure. The third panel shows the point-maximised metric trace, and the last panel shows the point-minimised metric trace (at $M_1$ in both cases, and constrained to the same equipotential surface of the quantum weight). We see that we have very strong qualitative freedom in changing the shape of the metric trace even without changing its mean value, and that the approximate threefold symmetry in the hoppings is insufficient to protect a threefold symmetry in the geometric response. 

\begin{figure}
    \centering
    \def\svgwidth{.95\linewidth}
    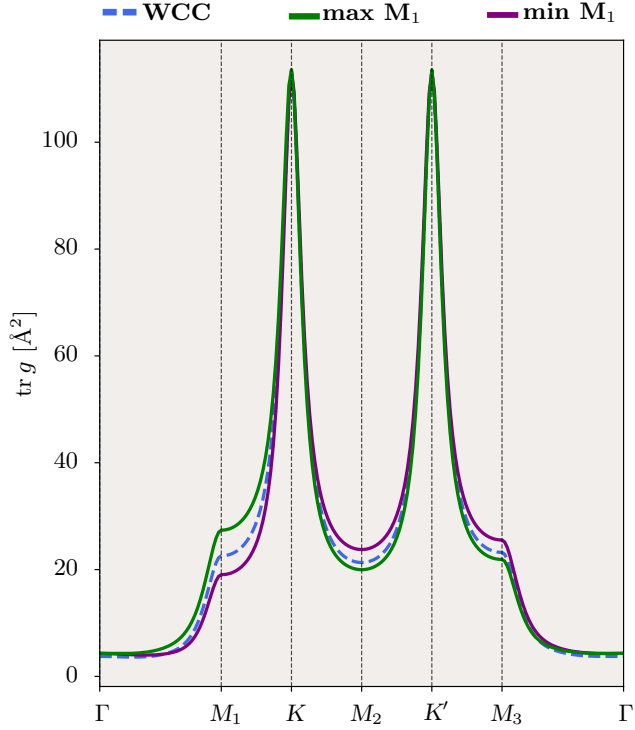
    \caption{Metric trace for V$_2$O$_3$ along a path in the BZ (see Fig.~\ref{fig:V2O3}), for WCCs (blue) as well as for embeddings maximising (green) and minimising (purple) the metric trace value at $M_1$, subject to the bound that no orbital may move more than $5\%$ in either reduced coordinate direction away from the WCCs.}
    \label{fig:V2O3_eps}
\end{figure}

These results are striking. %, indicating that geometric effects are very sensitive to embeddings. 
One might hope, however, that small perturbations, such as might occur from noise, should only quantitatively change the response. Indeed, by running an $M_1$-point maximisation and minimisation where we restrict the orbitals to move at most $5\%$ of a unit cell side length in each reduced coordinate direction, we see in Fig.~\ref{fig:V2O3_eps} that this holds for our example, though the quantitative change is not negligible. We also notice that, though this $M_1$-maximisation was not restricted to an equipotential surface of the quantum weight, the maximisation and minimisation is inverted away from $M_1$, indicating that the quantum weight is more stable to position perturbations than the local shape of the metric trace.

\subsection{Discussion\label{ssec:geo_discussion}}

\subsubsection{Engineering the quantum metric}

One reason why the quantum metric has received significant interest is because of its relationship to the quantum Fisher information, which signals where a system is sensitive to external responses. This is the idea underpinning critical quantum sensing, which employs this enhanced sensitivity for improved quantum sensors. This has been widely explored in a non-hermitian context, usually through exceptional-point sensing \cite{budich_non-hermitian_2020,chen_exceptional_2017,kartik_scaling_2025}, but it has also been explored in hermitian systems \cite{Sarkar2022Free-FermionicSensors}. A common problem in this approach is that the Fisher information usually is large wherever the gap is small, and consequently the system must operate slowly to ensure adiabaticity \cite{liu_experimental_2021}. This results in a trade-off between sensitivity and integration time.

We have shown that the shape of the metric trace can undergo drastic qualitative changes under a change of orbital embedding, even when keeping its average fixed. As such, there may be moderate band gap physical platforms which nevertheless exhibit a large metric locally. While we suggested flat plateaus around the minimal band gap as a promising place to look for interesting geometric effects, it remains interesting to make a more rigorous analysis looking for easily detectable markers of interesting geometry in physical platforms.

Furthermore, since the orbital degrees of freedom allow the size and shape of the metric to change without altering the band structure, there is a possibility (in principle) of \textit{engineering} systems with a very large metric even without a small gap. Such a system could avoid or diminish the trade-off between sensitivity and integration time in critical quantum sensing, and be used more generally to engineer interesting phases.

An important question to ask is then if it is possible to realize such an embedding change in a real experiment. We emphasize that our key previous results -- both that it is possible to obtain tight-binding models which reproduce desired geometric properties without sacrificing simplicity or band structure correctness, and that the metric depends qualitatively on embeddings -- do not depend on being able to physically implement these perturbations.

\subsubsection{Experimental feasibility of tuning geometry}
It is important to highlight that, while the position DOFs (orbital embeddings) and the energetic DOFs (hoppings) are completely independent in a tight-binding model, they are related physically to matrix elements of Wannier functions. In a physical platform, we may conceivably have some control over the Wannier functions, but changing these will generally alter both positional and energetic DOFs. The extent to which one is justified in treating these sectors as independent in real materials is unclear but could in principle be investigated experimentally. One conceivable way of doing this is to attempt to break a symmetry in the position sector independently and verify that symmetry-disallowed geometry independent responses remain zero while geometry dependent responses are not. This would give an indication that the view suggested by tight-binding of treating these sectors as independent is physically justified.

From the Wannier picture, it seems that if a symmetry is broken in the Wannier functions (giving symmetry breaking in the position sector), then the energy sector would have to be completely fine-tuned to retain that symmetry. However, one could look for a perturbation that only alters one section. Since the hoppings are typically considered more sensitive than the effective positions in analytical models (under small strain, for example, positions change linearly with strain whereas hoppings change exponentially \cite{botello-mendez_toward_2018}), the converse perturbation (changing the hoppings without changing positions) could presumably be possible to some approximation. If this intuition holds, symmetries in the energy sector could conceivably be independently broken. To break symmetries independently in the position sector seems more difficult, though one could (in principle) imagine taking a system out of symmetry and using a fine tuned energy perturbation to restore this symmetry only in the energy sector, while retaining the symmetry breaking in the position sector.

Going further, the option of directly controlling the orbital embeddings experimentally is highly attractive. However, while it is theoretically conceivable to fine-tune the Wannier functions so that hopping elements remain approximately fixed while their effective centres move, such a situation seems highly contrived physically. It may be achievable in some metamaterials \footnote{Less interestingly, sector specific perturbations should be achievable in analogue simulators.}. Furthermore, in Slater-Koster tight-binding \cite{slater_simplified_1954}, hopping elements depend on orbital orientation in addition to positions. When the hoppings are determined simultaneously by positions and other parameters like rotations, one could in theory imagine judicious combinations of translations and rotations which keep the hoppings constant, though it is unclear whether such fine-tuned combinations would be more than modelling artifacts.

\subsubsection{Symmetry-projected geometry dependence\label{sssec:sym_proj}}

It is generally true that, if there are fixed maximal Wyckoff positions for a model, then the minimal quantum weight will be attained at these \cite{huhtinen_revisiting_2022}. However, with free Wyckoff parameters, there is no obvious choice for embeddings, and the minimal quantum weight position will therefore not generally coincide with the minimal peak position. This underscores the importance of fixing these DOFs in a principled way using physical observables. It is furthermore important to realise that the MLWF WCCs do \textit{not} correspond to the minimal quantum weight embeddings in practice, both due to symmetry breaking and due to the point above.

A situation without a unique symmetry position is illustrated for a toy model in Fig.~\ref{fig:QWZ_assym} (App.~\ref{app:suppl_figs}), where the peak value of the metric trace changes by more than $100\%$ around an equipotential curve of the quantum weight. Notice therefore that the dramatic effects of the geometry perturbation does not rely on symmetry breaking, though a simple symmetric energy sector like in Figs. \ref{fig:Haldane} or \ref{fig:QWZ} makes the response to orbital displacements more easily predictable. 

The geometry dependence of such asymmetric two-band models is of some interests. Particularly, while this paper strictly considers non-symmetry constrained optimisations, it would be interesting in future work to explore whether the two band asymmetrised model's geometry dependence, in Fig.~\ref{fig:QWZ_assym}(b)--(c), is similar to a real multi-band system after projecting to the symmetry allowed region in the position sector. Concretely, one would then make a symmetry adapted Wannierisation of a material with two free Wyckoff parameters $(x_1, x_2)$ and make an analogous plot to Fig.~\ref{fig:QWZ_assym}(b)--(c) for these parameters. Two particularly interesting questions are: (i) Is there any interpretable latent structure in the symmetry-allowed region which could further suggest physically correct positions? In particular, we see the imprint of the symmetry group in the peak plots for Haldane and QWZ and the clear radial increase in the metric trace with orbital separation. Symmetry-projection should remove obvious structure, but certain patterns may remain in different responses, analogous to the triangle in the metric trace peak plot for the asymmetric QWZ variant. (ii) Is the freedom after symmetry projection as dramatic as that in Fig.~\ref{fig:QWZ_assym}? Does the hope that the optimal embeddings typically live inside the symmetry-allowed region, so that symmetry projection serves to optimise the search for optimal embeddings rather than reducing the ability to obtain good fits, hold for real materials?

\section{Conclusion \& Outlook\label{sec:conclusion}}
\subsection{Conclusion}
In this work, we have explored the role that orbital embeddings play in tight-binding models. We have shown that the ability of tight binding to capture geometry dependent observables, including the quantum geometric tensor as well as linear and non-linear optical responses, depend crucially on the choice of orbital embeddings, and have derived simple expressions describing geometry dependence. Using these expressions, we presented a concrete, efficient recipe for constructing tight-binding models which reproduce selected geometric effects at essentially DFT-level accuracy without sacrificing simplicity or band structure correctness, proceeding through a careful tuning of orbital positions. In particular, the results  challenge the expectation that tight-binding models are generally unsuited to capture non-linear responses \cite{wang_first-principles_2017, ibanez-azpiroz_ab_2018, ibanez-azpiroz_assessing_2022, ghosh_choosing_2025}. We further demonstrated the sensitivity of tight-binding models to the choice of orbital embedding, producing dramatic qualitative adjustments to the shape of the quantum metric under different embeddings.

In a broader sense, our results suggest a novel way to think about the role of orbital embeddings in tight-binding models. The diagonal tight-binding approximation inherently entails a choice of \textit{which} elements of the full position operator to retain. In conventional schemes, this is usually chosen either for simplicity (e.g. placing the orbitals on the atoms or at the origin), for symmetry (constraining the orbitals to certain Wyckoff positions) or to reduce the range of relevant hoppings (maximally localizing Wannier functions). These choices will often disagree, be arbitrary or fail to uniquely determine the embeddings. It is therefore attractive to use actual physical observables to fix these parameters. Still, it remains true in this approach that the correct choice of orbital embeddings depends at least to some extent on the relevant application of interest for the model, so that a trade-off between generality and accuracy persists.

\subsection{Outlook}
The most direct application of our results is to make optimally embedded tight-binding models for specific responses and materials. This would facilitate the study of geometric quantities in toy models, and enable better quantitative predictions. 

It would further be interesting to explore such symmetry constrained optimal embeddings. Indeed, a potentially undesirable property with optimal embeddings is that they may not respect the symmetries of the Bloch Hamiltonian. However, optimal embedding and symmetry are not necessarily in conflict, as there exist many symmetry groups where the relevant Wyckoff positions have free parameters, particularly for models of at least moderate size. In such cases, geometric observables can be analysed in terms of the Wyckoff parameters, and these may be fixed so as to best reproduce a geometric property of interest. This was discussed in some detail in Sec.~\ref{sssec:symmetry} with further applications mentioned in Sec.~\ref{sssec:sym_proj}.

An additional avenue to explore is the relationship between different modelling regimes and embeddings. For example, \cite{balut_fundamental_2026} attributes discrepancies between experimental measurements of the quantum weight and an empirical tight-binding model to correlation effects. As we have shown, however, the orbital embedding can drastically change the quantum weight. This suggests an interesting future research direction, where one quantitatively compares which experimental discrepancies arise due to correlation effects, and which arise due to the choice of orbital embedding in the tight-binding model.

It is also interesting to explore how sensitive various quantities are to orbital positions. In Sec.~\ref{sssec:band_gap_dep} we discussed likely reason why some regions in frequency space are more sensitive to geometry than others, though a more thorough analysis would be welcome. 
A rough estimate of the sensitivity of a response to the geometric embedding is the polynomial degree of the Berry connection $A$ which appears in the response. Still, this cannot be the full story, as certain quantities that depend on $A$ in a non-trivial manner are nevertheless independent of geometry (most notably the Chern number \cite{simon_contrasting_2020}). It would be interesting to investigate whether other quantized responses, such as the quantized circular photogalvanic effect \cite{de_juan_quantized_2017} or the quantised integrated shift effect \cite{Wojciech_quantized} are similarly geometry independent, at least given orbital displacements that do not break symmetries on which the quantisation relies. Similarly, it would also be interesting to investigate how changing the orbital positions can affect other model parameters frequently derived from tight-binding models, such as e.g. magnetic exchange couplings. More generally, finding properties (also beyond optical responses) that are geometry independent, even if they are not quantised, remains largely unexplored.

The important question of whether moving the orbital positions without modifying the hoppings can be thought of as a practically realisable effect or only as an artifact of the modelling, i.e. whether we can engineer orbital embeddings, remains open. We argue that this is likely difficult in practice, though we discuss some potential routes to realising this in Sec.~\ref{ssec:geo_discussion}. Understanding whether orbital engineering is possible is particularly interesting in light of our results on artificially engineering the quantum metric, presented in Sec.~\ref{sec:geometry}, which suggest that control over the embeddings would allow one to achieve interesting and surprising phases, including phases where both the quantum metric and the band gap are large. Either way, we reiterate that optimally embedded tight-binding remains useful for improving minimal models independent of physical realization.

Finally, we note that recent work has attributed physical significance to the minimal orbital embeddings in superconductivity. Indeed, changes in orbital positions dramatically affect the superfluid weight as they enter through the Peierls coupling. In \cite{huhtinen_revisiting_2022}, it is argued that the physically correct thermodynamic response is obtained by choosing the orbital positions that \textit{minimize} the metric trace. One might imagine similar constraints, imposed e.g. by optical sum rules \cite{cardenas-castillo_detecting_2024}, which attribute physical significance to certain orbital configurations.

More broadly, our results emphasize the well-known, but often under-appreciated, fact that significant care must be taken when predicting geometry-dependent responses from Wannierised models, or other tight-binding models, even if they reproduce experimental band structures well. This applies to a range of properties, also beyond those investigated here.

\textit{Note added:} During the finalization of our manuscript, \cite{Murakami_embedding} appeared, which independently suggests geometric engineering and discusses orbital embeddings in linear optical responses of a toy model, and \cite{Simon_embedding} appeared which discusses the interplay of minimal orbital embeddings and symmetries.

\section*{Acknowledgments}
We thank Wojciech J. Jankowski for helpful discussions, particularly concerning the physical interpretation of orbital embeddings, and Carl Fredrik N. Knutsen for insightful discussions. The simulations were performed on resources provided by Sigma2 -- the National Infrastructure for High-Performance Computing and Data Storage in Norway. G. F. L. acknowledges funding from the European Union’s Horizon Europe research and innovation programme under the Marie Skłodowska-Curie Grant Agreement No. 101126636, and the YoungCAS fellowship ENQUIRY, awarded by the Centre for Advanced Studies at the Norwegian Academy of Science and Letters.

\appendix

\section{Computational details\label{app:computation}}

\subsection{DFT and Wannierisation}

First-principle calculations were performed within the framework of plane-wave density functional theory (DFT) as implemented in the \texttt{QuantumESPRESSO} package \cite{giannozzi_advanced_2017, giannozzi_quantum_2009} and \texttt{Wannier90} was used for Wannierisation \cite{mostofi_updated_2014}. Note that the accuracy of the ab-initio computations is of only secondary importance for our purposes, as we are primarily interested in comparing differently embedded tight-binding models against DFT (taken as ground truth). As such, we are justified in making minor approximations which we do not expect to qualitatively change results, such as using known lattice parameters without explicit relaxation, neglecting weak SOC and omitting scissor corrections. Qualitative agreement is ensured by general agreement with band structures from previous literature, as well as reproducing the known shift current spectrum of GaAs \cite{ibanez-azpiroz_ab_2018} and the Chern topology of V$_2$O$_3$ \cite{mellaerts_two_2021}. 

For GaAs, we run DFT and Wannierisation following tutorial 25 of \texttt{Wannier90} \cite{mostofi_updated_2014, ibanez-azpiroz_ab_2018}. GaAs has a zincblende structure with lattice constant $a = 5.65$\,Å. We used energy cutoffs of 50\,Ry and 200\,Ry for the wavefunction and charge density respectively and an $8\times8\times8$ $\kv$-point grid \cite{monkhorst_special_1976} for the self-consistent field (SCF) calculation. 
Exchange and correlation effects were treated within the generalized gradient approximation (GGA) using the Perdew-Burke-Ernzerhof (PBE) functional \cite{perdew_generalized_1996}. The interaction between valence electrons and ionic cores was described using scalar relativistic projector augmented-wave (PAW) pseudopotentials \cite{kresse_ultrasoft_1999} from the PSLibrary \cite{dal_corso_pseudopotentials_2014}. The calculations were performed without spin polarization, corresponding to a non-magnetic bulk ground state. Wannierisation proceeded with the $s$ and $p$ orbitals of Ga.

For CdS with wurtzite structure, we used the lattice parameters $a = 4.17$\,Å and $c = 6.78$\,Å from the Materials Project for CdS (mp-672) \cite{horton_accelerated_2025, noauthor_materials_2020}. We used energy cutoffs of $60$\,Ry and $600$\,Ry for the wavefunction \cite{biswas_electronic_2022} and charge-density respectively and an $8\times8\times 6$ $\kv$-point grid for the SCF calculation. Exchange and correlation effects were again treated with GGA using the PBE functional, and interactions between valence electrons and ionic cores were again described using scalar relativistic PAW pseudopotentials from the PSLibrary without spin polarization. Wannierisation proceeded by projecting onto s and p orbitals of Cd and p orbitals of S. The band structure of the resulting tight-binding model is shown in Fig.~\ref{fig:CdS_response}(g), in excellent agreement with previous work \cite{chang_electronic_1983}, and with a similarly underestimated gap compared to experimental values. We emphasize that the approximations we make (neglecting SOC, skipping structural relaxation and scissor correction) are justified since our goal is to explore geometry dependence in tight-binding against a DFT baseline. Indeed, previous work without these approximations show low frequency optical responses of CdS \cite{nastos_optical_2010} seemingly in disagreement with our DFT-interpolated reference responses, indicating that second order optical responses may be sensitive to these approximations. We could not find their band structure for comparison.

As usual, disentanglement breaks symmetry slightly, both for GaAs and CdS, so that symmetry disallowed components of the Wannier interpolated optical responses are not exactly zero. Still, they are orders of magnitude smaller than the symmetry-allowed responses, as shown in Fig.~\ref{fig:GaAsSC}.

The Chern insulator V$_2$O$_3$ is a 2D material with a planar honeycomb-kagome structure \cite{mellaerts_two_2021}, which we studied using a periodic supercell geometry with a vacuum spacing of $17.0$\,Å. After relaxation with a force convergence criteria of $10^{-4}$\,Ry/Bohr, we find lattice parameters of $a =6.26$\,Å. We used energy cutoffs of 90\,Ry and 720\,Ry for the wavefunction and charge densities respectively, and a $12\times12\times1$ $\kv$-point grid. Previous studies have found that the energy gap in V$_2$O$_3$ strongly depends on the Hubbard correction $U$ \cite{mellaerts_two_2021}. As such, we treat exchange and correlation effects within GGA+U using the PBE functional with fully relativistic PAW pseudopotentials. We used the rotationally invariant formulation of DFT+U \cite{dudarev_electron-energy-loss_1998}, with an effective Hubbard parameter $U=3.28$\,eV \cite{mellaerts_two_2021} applied to the V (3d) orbitals. Spin-polarized calculations were initialized with a finite magnetic moment on the V atoms. Spin-orbit coupling was explicitly included through fully relativistic non-collinear calculations, yet the ground state was found to be a collinear ferromagnet with moments out-of-plane and a total magnetic moment of $4.01$\,$\mu_B$ per unit cell. Wannierisation was performed using the d-orbitals of V. By integrating the Berry curvature with \texttt{WannierBerri} \cite{tsirkin_high_2021}, we confirm the expected $C = 1$ topology for an isolated group of valence bands.

\subsection{Fitting optical responses\label{subapp:optical}}

We use \texttt{WannierBerri} to compute smooth high resolution plots for optical responses for GaAs and CdS. We use a frequency spacing of $0.03$ eV. % check units
For linear optical conductivity, a $\kv$-grid of $85^3$ was used with a fixed smearing of $0.1$. For second order responses, a $120^3$ grid was used and a smearing of $0.05$. For the shift current prinicipal value smearing, we used $\eta = 0.04$\,eV as in \cite{ibanez-azpiroz_ab_2018}.

Compared to postprocessing in Wannier90, which gives near identical plots as those in \cite{nastos_optical_2006, ibanez-azpiroz_ab_2018}, \texttt{WannierBerri} is known to smooth out the sharpest edges in the optical response with default settings. Energy smearing will happen in practice in experiments and the degree is mostly a matter of convention. 

\begin{figure*}[p]  % or [!p] to push it to a float page
  \begin{minipage}{0.48\textwidth}
    \centering
    \def\svgwidth{\textwidth}%
    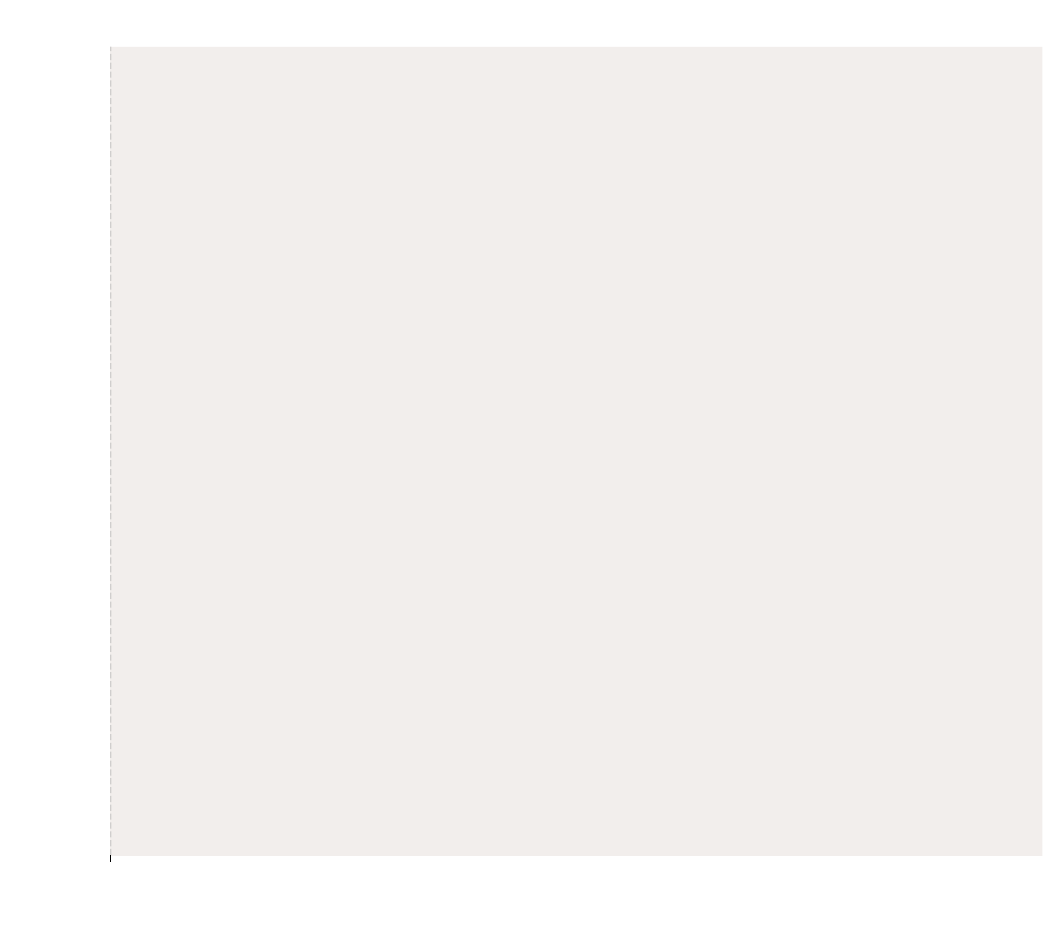
    \caption{Comparison of GaAs $xyz$-shift current response as computed with \texttt{postw90} \cite{mostofi_updated_2014} compared with \texttt{WannierBerri} (WB) \cite{tsirkin_high_2021}. WannierBerri with default settings gives a smoother curve than postw90, but the results generally agree closely. All responses in this work use WannierBerri, so this difference does not affect the comparisons presented between ab-initio and differently embedded tight binding models.}
    \label{fig:wb_vs_w90}
  \end{minipage}\hfill
  \begin{minipage}{0.48\textwidth}
    \def\svgwidth{\textwidth}%
    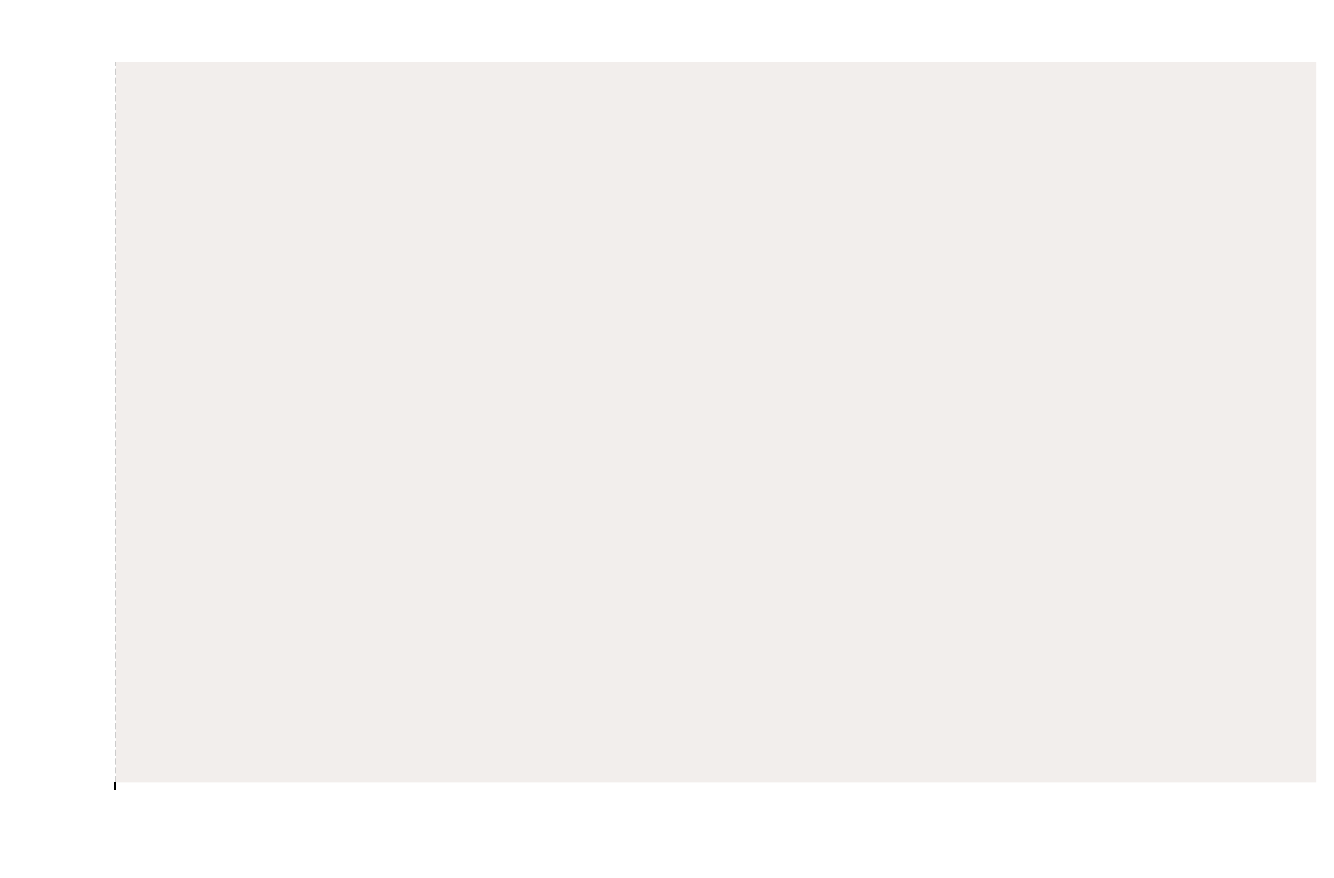
    \caption{We fit the optimal embeddings for optical responses on a very coarse grid, where responses oscillate wildly. Still, the fits seem to generalise well to the dense case as seen in Fig.~\ref{fig:GaAsSC}. The dashed line marked 'dense' is the true response computed on a dense grid with WannierBerri. The dotted response marked REF is the response computed with WannierBerri on a coarse $\kv$-grid, which agrees exactly with our computation (marked 'target', green) which we fit the Shift Current Tensor embeddings (SCC, orange) against. The blue line marked WCC is the original response before fitting, obtained from the MLWF WCCs on the coarse grid.}
    \label{fig:coarse_fit}
  \end{minipage}
\end{figure*}

Beyond the smoothing, the curves are in excellent agreement, as seen in Fig.~\ref{fig:wb_vs_w90}. Since all responses in this paper are computed with \texttt{WannierBerri}, the smoothing makes no difference to our conclusions regarding comparisons of responses. 

For obtaining optimal embeddings, we use a coarse $\q$-grid ($15^3$ for GaAs and $15^2\times12$ for CdS). For frequency resolved fits, we implement the formulae in section \ref{sssec:optical_responses}, treating delta functions as in \texttt{WannierBerri}. Indeed, we find that our computation agrees with \texttt{WannierBerri} for the minimal $\q$-grids, as seen in Fig.~\ref{fig:coarse_fit}. Notice in the figure that the optical response curves computed on such a coarse grid oscillate wildly and hence do not resemble the smooth curves obtained from a dense grid computation. Nevertheless, the embeddings obtained from fitting the coarse response seem to generalise well to the dense case. We ran the fits using the \texttt{trust-exact} solver from \texttt{SciPy} \cite{virtanen_scipy_2020}.

\subsection{\label{subapp:num_geo}Qualitative effects of geometry} 

For $2$D two-band toy models, it is a simple matter to compute relevant observables on a 2D grid corresponding to all orbital separation vectors. In testing different parameters for QWZ and Haldane models, we find indications that as flat a plateau as possible in the band structure is ideal for large relative changes in peak metric trace, in agreement with the rough expectation in section \ref{sssec:band_gap_dep}. As such, we consider a Chern phase Haldane model with hopping parameters $\Delta=0.3$, $t_1=-1$ and $t_2=-0.21$ (with a Peierls phase of $\pi/2$ as is usual), with band structure shown in Fig.~\ref{fig:Haldane}(a). We vary the embeddings on an equipotential curve of the quantum weight. Specifically, we use the value of the quantum weight corresponding to the geometry independent contribution obtained by placing both orbitals at the origin. The minimisation and maximisation in Fig.~\ref{fig:Haldane}(b)--(c) are hence made on this equipotential curve, which is marked by a black dotted circle in the panels.

In appendix \ref{app:bounds}, we also consider a Chern phase QWZ model with KPC Hamiltonian specified by $\vec{d} = (-\sin k_y, - \sin k_x, u +\cos k_x + \cos k_y)$ with $u = 1$, with a bound arbitrarily chosen to be the quantum weight at orbital separation $(1/2, 1/2)$.  We also consider an asymmetrised QWZ variant with KPC Hamiltonian specified by $\vec{d} = (-\sin k_y, - \sin k_x + \d \cos k_y, u +\cos k_x + \cos k_y)$ with $u = 1$ and $\delta = 1/4$.

To investigate the effect of the position sector on the metric trace in a real material, we consider monolayer V$_2$O$_3$. We pick a moderate band gap $M$-point, with reduced coordinate $\kv = (1/2, 1/2)$, and maximize and minimize the metric trace at this point for positions on a high dimensional equipotential surface of the quantum weight. We obtain a rough bounding value by sampling positions uniformly in reduced coordinates with $\tau^\a_i \in [-1,1]$ $1000$ times, computing the quantum weight for each sample and taking the average. 

This value then defines a high-dimensional equipotential surface. Clearly, we cannot easily plot it as we did for the $1$-dimensional equipotential curve in Fig.~\ref{fig:Haldane}(b). However, moving the positions inside this high-dimensional equipotential surface to maximise the peak at $M_1$ should be visualised in the same way as the familiar picture of e.g. a multi-particle system evolving in configuration space on an equipotential surface of constant energy to settle into a state maximising the entropy.

To run the constrained optimization, we used the \texttt{trust-constr} solver from \texttt{SciPy} \cite{conn_trust_2000, virtanen_scipy_2020} with the non-linear constraint that the quantum weight should be less than the bounding value. We provide the Jacobian and Hessian for both the cost and constraint, computed as described in Sec.~\ref{sec:theory} and App.~\ref{app:qgt}. By using the same optimizer with bounds, allowing only a displacement of $5\%$ of a unit cell side length in each reduced coordinate direction relative to the MLWF WCCs, we can also investigate the effect of small position perturbations as shown in Fig.~\ref{fig:V2O3_eps}.

\section{\label{app:qgt}Fitting the quantum geometric tensor}

To find optimal embeddings quickly for general geometric observables, or to run efficient minimisation or maximisation of tight-binding quantities like the metric trace, we want to take position derivatives to obtain the Jacobian and Hessian. This is a simple matter of taking derivatives of the geometry dependent term from the decomposition in Sec.~\ref{ssec:decomposition}, and we show the procedure in some detail for the QGT here.

\subsection{Position derivatives}
The computations below are simple but index intensive. While great for numerical implementation, it is not particularly readable, so it is worth discussing the notation explicitly.  Matrices are denoted by non-italic capital letters and can depend on the band index (labeled by $n,m$) such as the projectors $\Pm_n$, or on periodic direction (labeled by $i,j,k,\ell$) such as the derivatives $\ptl_j \Hm$ or diagonal position matrices $\D_j$. These indices will always be explicit. 

However, it will also be convenient to access the orbital basis components of matrices, such as $P_n^{\a\b}$. We use greek letters $\a,\b, \g, \eta$ for this. We will use early alphabet letters $a, b$ instead of greek letters to indicate when an orbital index is implicitly summed over. The capitalised trace $\Tr$ acts on matrices, that is, on these orbital indices. 

Consider two disjoint groups of bands, such as the occupied and unoccupied bands of an insulator, and restrict the index $n$ to the first group and $m$ to the latter. Note in particular that this implies always $E_n \neq E_m$. In practice, we compute the optical QGT by Kubo-expansion, using Eq.~\eqref{eq:A_Kubo},
\begin{align*}
    \qgt_{ij}^{nm} \equiv \braket{\ptl_i u_n|\Pm_m| \ptl_j u_n} = \frac{\Tr\, [\Pm_n \partial_i \Hm \:\Pm_m \partial_j \Hm]}{(E_n - E_m)^2}.
\end{align*}
Similarly, we can expand the displacement in Eq.~\eqref{eq:dQ_opt} as
\begin{align*}
    \d \qgt_{ij}^{nm} =& \: \Tr\, [\Pm_n \D_i \Pm_m \D_j] \\ &-  i \Tr\, [ \D_i \frac{\Pm_m \partial_j \Hm \: \Pm_n}{E_m - E_n} + \frac{\Pm_n \partial_i \Hm \: \Pm_m}{E_n - E_m} \D_j]  \\
    =& \:P_n^{ab} \tau_i^b P_m^{ba} \tau_j^a  + i(\overset{*}{\mathrm{a}}{}_{j,a}^{nm} \tau_i^a - \mathrm{a}_{i,a}^{nm} \tau_j^a)
\end{align*}
where we introduced
\[
   \mathrm{a}_{j,\a}^{nm} \equiv [\frac{\Pm_n \partial_j \Hm\: \Pm_m}{E_n - E_m}]_{\a\a} = - [\frac{\Pm_m \partial_j \Hm \: \Pm_n}{E_m - E_n}]^\dagger_{\a\a} = - \overset{*}{\mathrm{a}}{}^{mn}_{j,\a}.
\]
The second equality used Hermiticity. Note that $\mathrm{a}_{j,\a}^{nm}$ can be precomputed.

Taking derivatives is easy, since $\partial_k^\gamma \tau_j^{\alpha} =  \delta^{\alpha\gamma} \delta_{jk}$. We get the Jacobian
\begin{align}
     \partial_k^\gamma &\delta \qgt^{nm}_{ij} =\\
     &P_m^{\g a} \tau^a_j P_n^{a\g} \delta_{ik} + P_n^{\g b} \tau_i^b P_m^{b\g} \delta_{jk} + i[\overset{*}{\mathrm{a}}{}_{j,\g}^{nm} \delta_{ik} - \mathrm{a}_{i, \gamma}^{nm} \delta_{jk}] \notag
\end{align}
and the constant Hessian
\[
    \partial_\ell^\eta \partial_k^\gamma \delta \qgt^{nm}_{ij} = P_n^{\eta\gamma} P_m^{\gamma\eta}  \delta_{ik} \delta_{j\ell} + P_n^{\gamma\eta} P_m^{\eta\gamma}  \delta_{i\ell} \delta_{jk}.
\]
The corresponding derivatives of the Abelian QGT are easily obtained by summing over $n$ and $m$.

Given a reference value $\qgt^{\mathrm{ref}}_{ij}$ for the QGT it is now straightforward to run a MSE-minimisation of \mbox{$\| \qgt^{\mathrm{ref}}_{ij} - \qgt_{ij}(\tau^\a_k)\|$,} using e.g. minimisers from \texttt{SciPy} \cite{virtanen_scipy_2020}. This works just the same for other geometric observables, including optical responses. For direct minimisation, like finding the minimial metric trace embeddings or tightening $\kv$-local bounds between tight-binding observables, no reference value is needed.

\subsection{The QGT from DFT}

As an aside, we note that for plane-wave DFT such as implemented in \texttt{VASP} \cite{VASP}, the single-particle wave functions are given as $\ket{\psi_n} = \sum_{\Gv} C_{n \Gv} \ket{\kv + \Gv}$, where the coefficients $C_{n\Gv}$ can be retrieved from the \texttt{WAVECAR}, e.g. with the Python package \texttt{VaspBandUnfolding} \cite{zheng_vaspbandunfolding_2017}. If SOC is neglected, the velocity operator becomes $\hat v_j = \hat p_j /m_e$, hence has plane wave eigenstates, and the optical QGT can be computed without coarse finite difference approximations as
\begin{align*}
    &\qgt_{ij}^{nm} = \frac{\braket{\psi_n| \hbar \hat v_i | \psi_m}\braket{\psi_m | \hbar \hat v_j | \psi_n}}{(E_n-E_m)^2} \\
    &= \frac{\hbar^4}{m_e^2}
    \sum_{\Gv, \Gv'} \frac{C_{n\Gv}^* C_{m\Gv} (k_i + G_i) C_{m\Gv'}^* C_{n\Gv'}  (k_j + G'_j)}{(E_n-E_m)^2}. 
\end{align*}
Summing over $n$ occupied and $m$ unoccupied gives the Abelian QGT as usual. This was previously suggested in \cite{kim_direct_2025}.

\section{\label{app:interspace_berry}Solving for the interspace Berry connection}

In this appendix, we show explicitly how to optimize embeddings against the \textit{interspace} Berry connection. Let $\Um_{n;}$ ($\Um_{m;}$) denote the rectangular block of $\Um$ obtained by restricting the first index to the occupied (unoccupied) indices. In what follows, we use bold for Cartesian vectors to avoid carrying the $j$-subscript. Restricting the minimization to the occupied-unoccupied rectangular block of the Berry connection, we can write out the OD-error with some simple algebraic manipulation 
\begin{align*}
    &\| \Um^\dagger_{n;} \tilde{\bm{\mathbbm{a}}} \Um^{}_{;m} \| = \int_{\kv} \Tr [\Um^\dagger_{n;}\tilde{\bm{\mathbbm{a}}}\Um^{}_{;m}\Um^\dagger_{m;}\tilde{\bm{\mathbbm{a}}}^\dagger \Um^{}_{;n}] = \int_{\kv} \Tr[\tilde{\bm{\mathbbm{a}}}\Qm\tilde{\bm{\mathbbm{a}}}^\dagger \Pm] \\
    &=\int_{\kv}\left[ \:\:\sum_{\R,\R'} e^{i\kv \cdot (\R -\R')} \braket{\Ov a | \hat \rv | \R b} Q_{bc} \braket{\R' c | \hat \rv | \Ov d} P_{da} \right. \\
    &\left. - \left( \sum_{\R} e^{i\kv \cdot \R} \braket{\Ov a | \hat \rv | \R b}Q_{bc}\bm\tau_c P_{ca} + \mathrm{c.c.} \right) + \bm\tau_a Q_{ac} \bm\tau_c P_{ca}\right].
\end{align*}
Taking the derivative with respect to orbital $\a$, $\ptl/\ptl \bm\tau_\a$, the first term vanishes and we get the stationary criterion
\begin{align*}
    \Re \int_{\kv} \left([\Pm\tilde{\bm{\Am}} \Qm]_{\a\a} -\Pm_{\a b} \Qm_{b\a} \bm{\tau}_b\right) = 0.
\end{align*}
We define
\begin{align*}
    \mathcal{H}_{\a\b} = \int_{\kv} \Re(P_{\a\b}Q_{\b\a}) \quad&\text{and} \quad y_{\a j} = \int_{\kv}\Re(\Pm \tilde \Am_j \Qm)_{\a\a}
\end{align*}
and solve $\mathcal{H}_{\a b}\tau^b_j = y_{\a j}$ for $\tau^\b_j$ to get the optimal interspace embedding. 

While $\mathcal{H}$ is positive semidefinite (by the Schur product theorem), there is one zero eigenvalue corresponding to the uniform shift $(1,1,...,1)^\top$ (this is clear by orthogonality of the projectors $PQ = 0$, by applying $\mathcal{H}$ to $\bm \tau'_\a = \bm \tau_\a + \vec{c}$ for $\vec{c}$ independent of $\a$), so a pseudoinverse/least squares solution must be used.

\section{\label{app:MLWFs} MLWF WCCs minimize the projected space external Berry connection}
In this appendix, we prove that the MLWF WCCs choice minimizes the square Frobenius norm of the projected space external Berry connection $\mathbbm{a}$, given in Eq.~\eqref{eq:OD_error_H}. By unitary equivalence and Fourier normalisation, we have
\begin{align*}
\| \mathbbm{a} \| &= \| \tilde{\mathbbm{a}}\| =\sum_{\a\b}\sum_\R | \braket{\a \Ov|\hat \rv - \bm \tau_\a | \b\R} |^2 \\
=&\sum_\R \sum_{\a\b} |\braket{\a \Ov|\hat\rv|\b\R}|^2  - 2\bm\tau_\a\braket{\a\Ov |\hat \rv |\a\Ov} +\bm\tau_\a^2.
\end{align*}
The first term in the last line, carrying the sum, is independent of choice of positions, and therefore vanishes when differentiating. The stationary criterion is $\bm\tau_\a = \braket{\a\Ov|\hat\rv|\a\Ov}$, as expected.

Given this choice, we show further that the MLWF basis choice minimizes the remaining error
\[
    \sum_{\b\R \neq \a\Ov}|\braket{\a\Ov |\hat \rv |\b\R}|^2.
\]
The MLWFs correspond by definintion to the gauge choice minimizing the spread functional
\[
    \Omega = \sum_\a \big[ \braket{\a\Ov|\hat \rv^2 | \a\Ov} - \braket{\a\Ov|\hat \rv |\a \Ov}^2 \big].
\]
By adding and subtracting the off diagonal terms, we get the decomposition
\[
    \sum_\a\Big[ \braket{\hat \rv^2}_\a - \sum_{\b\R}|\braket{\b\R|\hat \rv|\a\Ov}|^2  + \sum_{\b\R \neq \a\Ov}|\braket{\b\R|\hat \rv|\a\Ov}|^2\Big]. 
\] 
The first two terms form a basis invariant object, namely the integrated trace of the Abelian metric for the projected space, hence is unaffected by basis choice, as shown in appendix C of \cite{marzari_maximally_1997}. The final term is precisely the TBA error given the WCC embedding. 

Note that this does \textit{not} show that the MLWF WCCs minimize the error on actual observables. In fact, this is not true, as discussed extensively in the text.

% ###

\section{\label{app:Chern}Geometry invariance of multiband Chern numbers}

In this appendix we prove that the multiband Chern numbers are geometry invariant. It is convenient to use projector notation. The Abelian Berry connection for an isolated band subspace with projector $\Pm$ is \cite{mitscherling_gauge-invariant_2025}:
\begin{equation} \label{eq:Om_proj}
    \Om_{ij} = -2 \Im \Tr \left\{ \Pm\ptl_i \Pm\ptl_j \Pm\right\}.
\end{equation}
An orbital displacement from primed to unprimed positions gives $\Pm = \Dm \Pm' \Dm^\dagger$ and expanding with the chain rule we find
\[
    \ptl_j \Pm = \Dm( \ptl_j \Pm' + i[\Pm', \D_j])\Dm^\dagger,
\]
where we commuted diagonal matrices $\Dm$ and $\D_j$ and used the standard bracket notation for the commutator of $\Pm'$ and $\D_j$. 
Plugging into \eqref{eq:Om_proj} and simplifying with cyclicity of the trace and $\Pm^2 = \Pm$ (from which it also follows that $\Pm (\ptl_j \Pm) \Pm = 0$), we find
\begin{align*}
    \delta\Om_{ij} &\equiv \Om_{ij} - \Om'_{ij} \\
    &= -2 \Im \Tr \{ (\D_i \Pm' \D_j + \mathrm{h.c.}) - (\Pm' \D_i \Pm' \D_j  + \mathrm{h.c.}) \\
    &\phantom{= -2\Im \Tr \{(} - i(\ptl_i \Pm' \D_j \Pm'   - \Pm' \D_i \ptl_j \Pm')\} \\
    &= 2\Re \Tr\{ \ptl_i \Pm' \D_j \Pm'   - \Pm' \D_i \ptl_j \Pm'\} \\
    &= \ptl_i \xi_j - \ptl_j \xi_i
\end{align*}
where we discarded the first two real terms in parenthesis to get the third line. This shows that the change $\delta\Omega_{ij}$ under embedding changes is an exact form, which vanishes by Stokes' theorem when integrating over the BZ. In the final line, we introduced $\xi_j = \Tr \{\Pm' \D_j \Pm'\}$ such that
\begin{align*}
    \ptl_i \xi_j = \Tr\{ \ptl_i\Pm' \D_j \Pm' + \mathrm{h.c.}\} &= 2 \Re\Tr\{\ptl_i \Pm' \D_j \Pm' \} \\
    &= 2 \Re\Tr\{\Pm' \D_j \ptl_i \Pm' \}.
\end{align*}

\section{\label{app:bounds}Bounds on geometry}
In this appendix we discuss various bounds on geometry, arising from topology and optics. Exploring these bounds are part of the motivation for Sec.~\ref{sec:geometry}.

\subsection{Topological lower bound}
There are well-known lower bounds on geometry in terms of topology \cite{Topology_bounds_PRL, bouhon2023quantumgeometryprojectivesingle,spin_bounds}. In 2D, we have a single Berry curvature component $\Om_{xy} = \Om$ and three metric components, which we denote by $g_x$, $g_y$ for diagonal components and $g_{xy}$ for the off-diagonal. We have well-known $\kv$-local bounds
\begin{equation} \label{eq:bounds}
    |\Om| \leq 2\sqrt{\det g} \leq 2\sqrt{g_{x}g_{y}} \leq \tr g,
\end{equation}
where the first inequality uses positive definiteness of the QGT,
\[
    0 \leq \det \qgt = \det g - \lb\frac{\Om}{2}\rb^2,
\]
and the last two follow from positivity of real squares
\[
    \det g = g_{x}g_{y} - g_{xy}^2 \leq g_{x} g_{y} \quad \text{and}\quad 0\leq (\sqrt{g_{x}}-\sqrt{g_{y}})^2.
\]

\begin{figure}[!t]
    \centering
    \def\svgwidth{1\linewidth}
    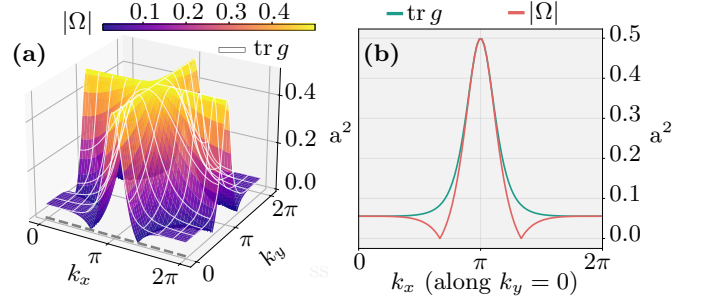
    \caption{\textbf{(a)} The metric trace (white mesh) and absolute value of the Berry curvature (surface) for the QWZ model with TBC Hamiltonian specified by $\vec{d}' = (-\sin k_y, - \sin k_x, u +\cos k_x + \cos k_y)$ with $u = 1$. The smooth metric trace cannot coincide with the non-smooth absolute value of the Berry curvature, as highlighted in \textbf{(b)} showing the cross section above $k_y=0$.}
    \label{fig:QWZ_bound}
\end{figure}

In the two-band case, we always have $\det \qgt = 0$ and so the first bound is trivial \cite{FCI_Flatness}. The other inequalities still need not be saturated. In fact, at any non-trivial zero of the Berry curvature, its absolute value is not smooth, and so the smooth metric trace cannot follow the motion of $|\Om|$. As such, we should generally not expect the final bound to saturate everywhere, at least when $\Om$ takes positive and negative values. This is illustrated in Fig.~\ref{fig:QWZ_bound}. 

One could hope that the orbital embeddings could be chosen to flatten $\Om$ to be of definite sign in a Chern phase, but this is not possible for any parameter choice in our QWZ model with KPC Hamiltonian specified by $\vec{d}' = (-\sin k_y, - \sin k_x, u +\cos k_x + \cos k_y)$ and orbital separation $(x,y) = (a, b)$. (This is seen e.g. by considering $\kv_- = (0,0)$ and $\kv_+ =(\pi,\pi)$ in the BZ where $d_x'=d_y'=0$: The triple product determining the sign of the Berry connection
\[
    \text{sgn } \Om' = \text{sgn } \left[ \dv' \cdot (\ptl_1 \dv' \times\ptl_2\dv') \right]
\]
 is computed to be $u\mp 2$ at $\kv_\pm$, which have opposite sign in the Chern phases $|u|\leq 2$. To include the orbital embeddings, notice that they act on $\dv'$ by a rotation $\mathrm{R}$ about the $z$-axis by an angle $\phi = ak_x + bk_y$, and that
 \[
 \ptl_i \dv = \ptl_i(\mathrm{R}\dv') =\ptl_i \mathrm{R} \dv' + R\ptl_i\dv'.
 \]
 At $d_x'=d_y'=0$, the first term vanishes and the triple product reads
 \[
    \mathrm{R}\dv' \cdot \mathrm R(\ptl_1 \dv' \times\ptl_2\dv') = \dv' \cdot (\ptl_1 \dv' \times\ptl_2\dv'),
 \]
 where we used that rotations distribute over cross products and preserve inner products. Hence, the signs remain opposite independent of embeddings, and we cannot saturate the trace bound for any parameter choice of the model.)

\begin{figure}
    \centering
    \def\svgwidth{1\linewidth}
    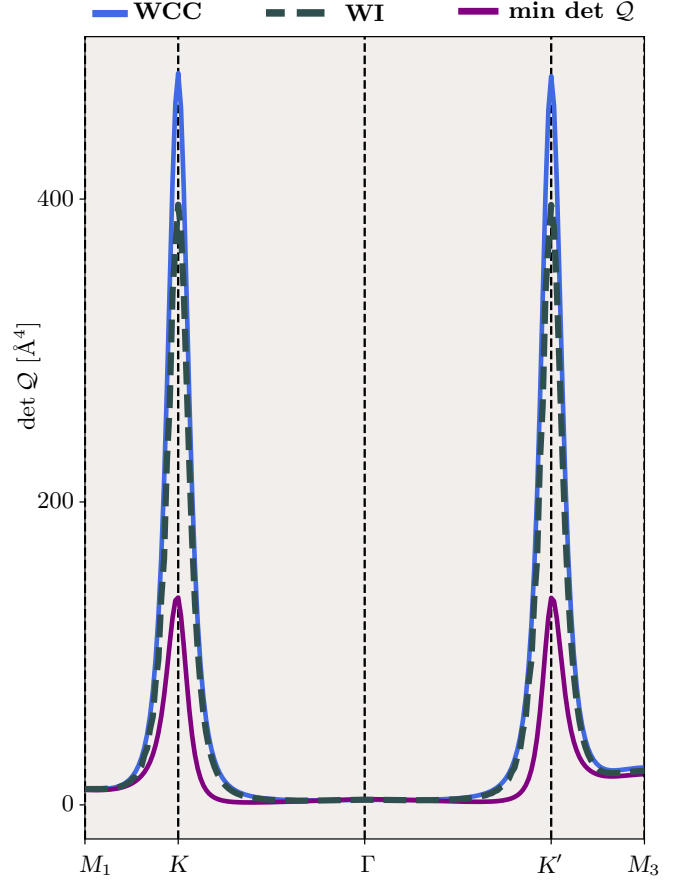
    \caption{Determinant of the QGT of a Chern subspace for V$_2$O$_3$ along a path in the BZ (see Fig.~\ref{fig:V2O3}), for WCCs (blue), Wannier Interpolation (WI) and the minimal determinant positions.}
    \label{fig:V2O3_detQ}
\end{figure}

In the multi-band case, the determinant bound is non-trivial. While the BZ average of the Berry curvature is geometry independent, so that any local increase caused by an orbital displacement must be compensated for elsewhere, the metric components have a larger freedom. Indeed, their average is strongly geometry dependent and is unbounded above, while its lower bound, corresponding to positions of minimal average metric trace, is believed to have some physical significance, entering expressions for superfluid weight \cite{huhtinen_revisiting_2022}. In light of this asymmetry in geometry dependence, one might ask to what extent we can artificially move the orbitals so as to bring the metric towards the Berry curvature. In Fig.~\ref{fig:V2O3_detQ}, we show that the determinant bound can be significantly tightened by tuning orbital embeddings for monolayer V$_2$O$_3$, and show that MLWF WCC tight-binding in this case somewhat overestimates the bound.

As a final note, as is well known \cite{onishi_quantum_2025}, we can use \eqref{eq:bounds} to get a lower bound for the quantum weight in terms of topology:
\begin{equation} \label{eq:opt_bound}
    2\pi|C| \leq \int_{\kv} \tr g = \frac{2\hbar}{e^2}\int_0^\infty \Re \frac{\tr\varsigma}{\om} \dd{\om}.
\end{equation}
Since topology is more accessible, both experimentally and theoretically, than geometry, this gives interesting lower bounds for several observables like superfluid weight and optical responses.

\subsection{Optical upper bound}

Finally, we provide for completeness a simple upper bound for the quantum weight, as in \cite{onishi_fundamental_2024}. 
By integrating out the delta-function in \eqref{eq:linear_response} and moving constants, one finds that the average metric trace satisfies
\begin{align} \label{eq:upper_bound}
    \int_\kv \tr g &= \frac{2\hbar}{e^2}\int_{\Om_g}^\infty \Re \frac{\tr\varsigma}{\om} \dd{\om}\\
    &\leq \frac{2\hbar}{e^2\Om_g}\int_{\Om_g}^\infty \Re \tr\varsigma \dd{\om} \equiv \frac{2\hbar I}{e^2\Om_g}\notag
\end{align}
where $\Om_g = \min \om_g(\kv)$ is the smallest band gap and we have introduced
\begin{equation} \label{eq:upper_bound_density}
    I = \int_{\Om_g}^\infty\Re \tr \varsigma \dd{\om} \leq \mathcal{I} = d\frac{\pi n e^2}{2m},
\end{equation}
where $\mathcal{I}$ is the same integral as $I$, except that the tight-binding optical conductivity is replaced by the total space conductivity with no maximal energy cut-off for intermediate states. The final equality then follows from a standard component-wise sum rule for the optical conductivity \cite{onishi_fundamental_2024}, where $d$ is the number of periodic dimensions, $m$ is the electron mass and $n$ is the total electron density (which one can upper bound by electron counting). 
Tighter bounds are possible in terms of experimental quantities like the permittivity and plasmon energy \cite{onishi_quantum_2025}.

%%%%%%%%%%%%%%%%%%%%%%%%%%%%%%%%%%
%%%%%%%%%% SUPPLEMENTARY FIGURES
%%%%%%%%%%%%%%%%%%%%%%%%%%%%%%%%%%%%%%%%%%

\clearpage
\onecolumngrid
\section{Supplementary Figures\label{app:suppl_figs}}
\vspace{50pt}

% =========================================================
% PAGE 1: QWZ & QWZ_asym Side-by-Side
% =========================================================
\begin{figure}[h!]
    \centering
    \begin{minipage}[t]{0.48\textwidth}
        \centering
        \def\svgwidth{\linewidth}
        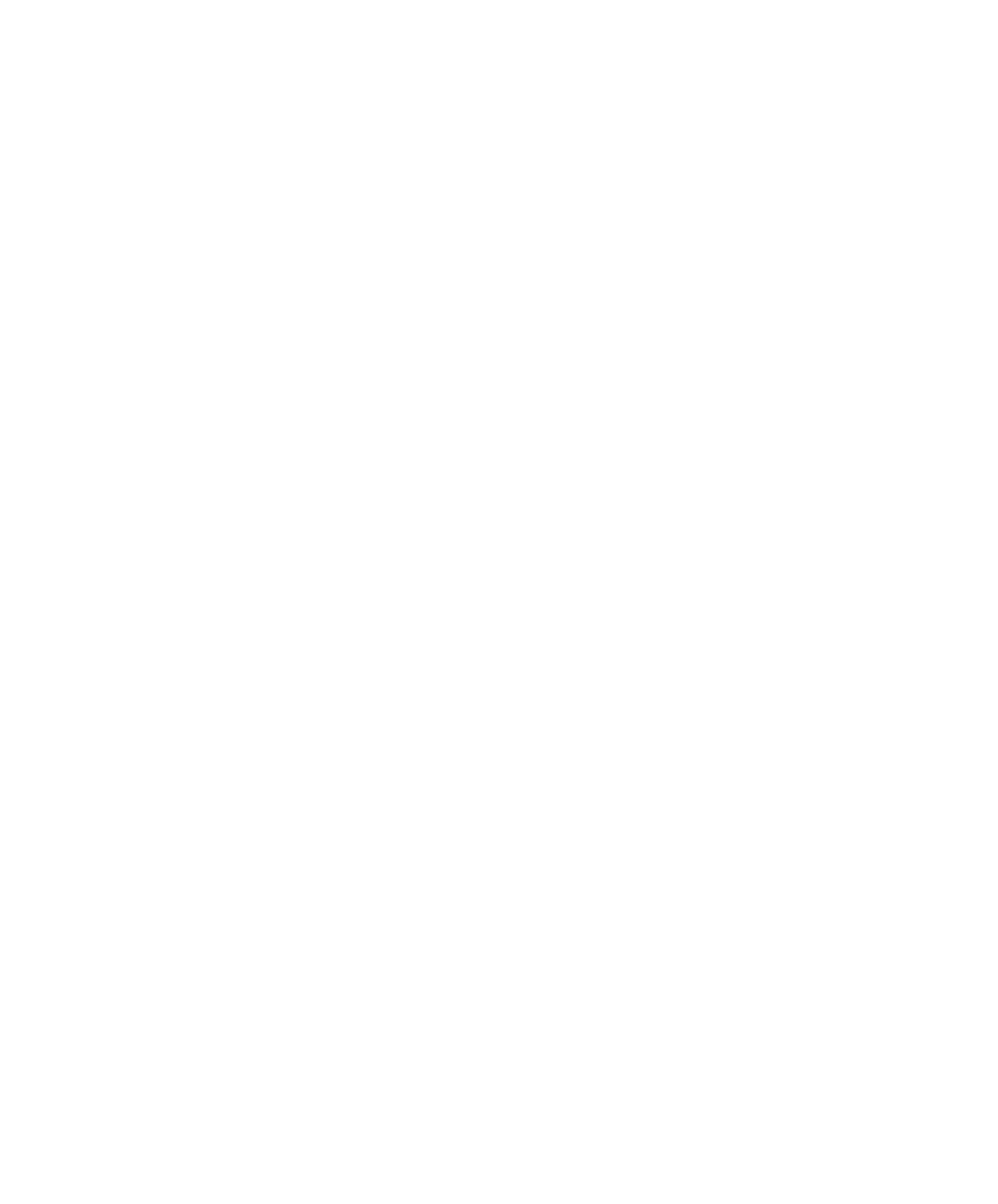
        \caption{Geometry dependence of the metric trace of the QWZ model with KPC Hamiltonian specified by $\vec{d} = (-\sin k_y, - \sin k_x, u +\cos k_x + \cos k_y)$ with $u = 1$. Please compare to Fig.~\ref{fig:Haldane}, as the setup is similar: \textbf{(a)} The band gap over the BZ. Notice the near-flat plus-shaped plateau. \textbf{(b)}--\textbf{(c)} The BZ average metric trace $\mathcal{K}$ as a function of orbital separation $(x,y)$, and the maximum of the metric trace $\mathcal{G}$ as a function of orbital separation, respectively. The unit cell is indicated, together with a (arbitrary) circular equipotential curve of the quantum weight. Three points are indicated, corresponding to plots (d)--(f) showing the metric trace in $\kv$-space: \textbf{(d)} The white triangle marks the maximal symmetry position, \textbf{(e)} the black dot marks the maximal peak position on the equipotential curve, which is one of four maximal peak positions, and \textbf{(f)} the red cross marks one of four minimal peak position on the equipotential curve.}
        \label{fig:QWZ}
    \end{minipage}%
    \hfill
    \begin{minipage}[t]{0.48\textwidth}
        \centering
        \def\svgwidth{\linewidth}
        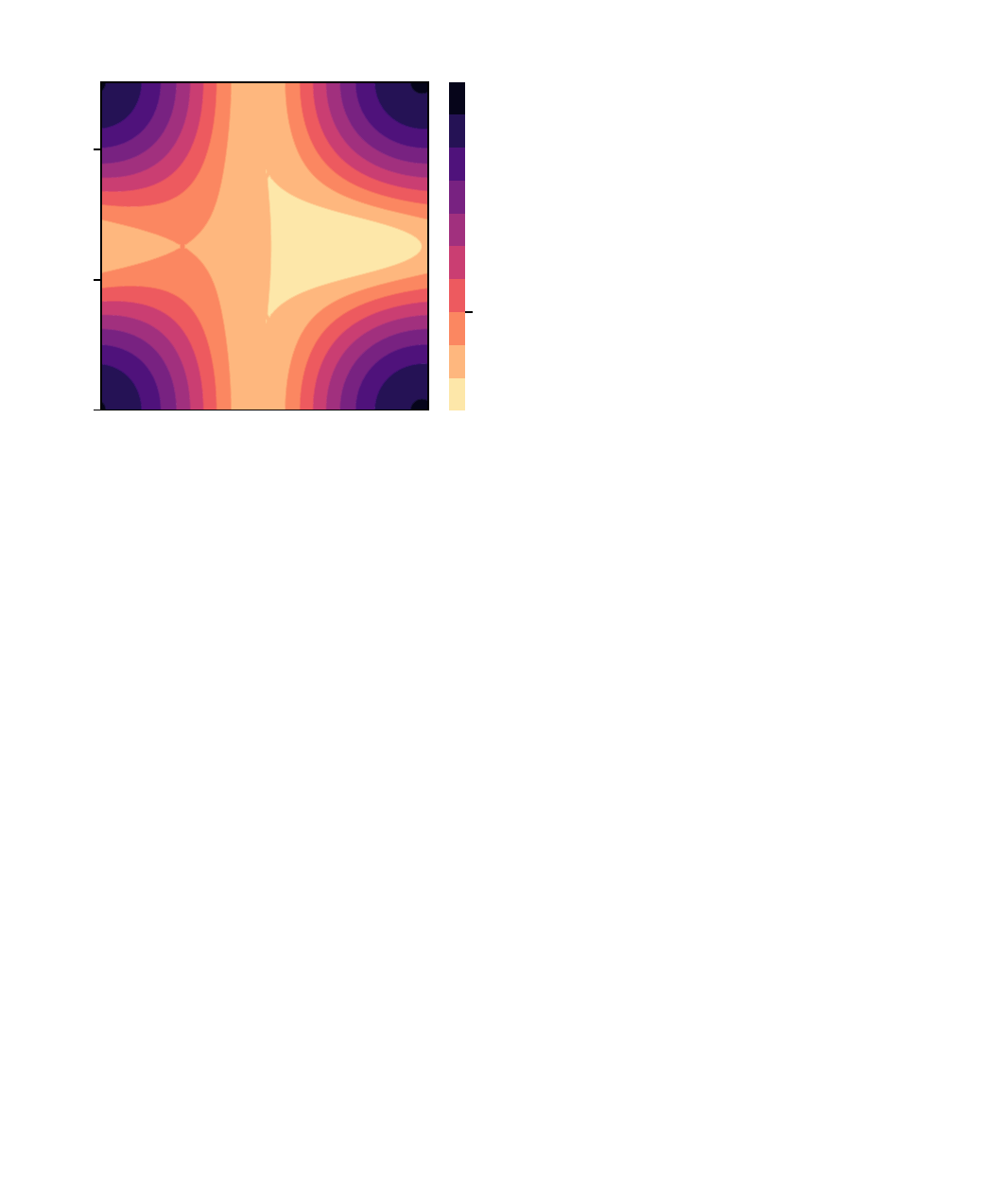
        \caption{Geometry dependence of the metric trace of an assymetric QWZ-variant defined by the KPC Hamiltonian specified by $\vec{d} = (-\sin k_y, - \sin k_x + \d \cos k_y, u +\cos k_x + \cos k_y)$ with $u = 1$ and $\delta = 1/4$. Please refer to Fig.~\ref{fig:QWZ} and the corresponding panel description, but notice that there is no maximal symmetry position in this case, so that the white triangle now specifically marks the minimal quantum weight position. The results are interpretable, as the hopping perturbation picks out a preferred direction (to the right in the BZ) which moves the minimal peak position to the embeddings which maximally resist this preference (pushing to the left in the BZ), while the maximal peak position is attained when the positions push in the same direction as preferred by the energy gap.}
        \label{fig:QWZ_assym}
    \end{minipage}
\end{figure}

\FloatBarrier
\twocolumngrid
\bibliography{references}

\end{document}